\documentclass[a4paper,11pt]{article}
\pdfoutput=1 % if your are submitting a pdflatex (i.e. if you 
\usepackage[T1]{fontenc} % if needed
\usepackage[dvipsnames]{xcolor}
\usepackage{slashed}
\usepackage{subcaption}
\usepackage{physics}
\usepackage{csquotes}
\usepackage{caption}
\usepackage{tikz}
\usepackage{tikz-cd}
\usepackage[compat=1.1.0]{tikz-feynman}
\usetikzlibrary{arrows}
\usepackage{jheppub} % for details on the use of the package, please
\usepackage{amsmath}
\usepackage{amssymb}
\usepackage{bbm}
\usepackage{hyperref}
\usepackage{dsfont}
\newcommand{\ignore}[1]{}
\usepackage{amsmath,bm}
\usepackage{framed}
\usepackage{multirow}
\usepackage[normalem]{ulem}

\newcommand{\Trh}{T_\text{rh}}
\title{\boldmath Modular invariant inflation and reheating}

\author[a]{Gui-Jun Ding}
\author[a]{Si-Yi Jiang}
\author[b,c]{Yong Xu}
\author[d,e]{Wenbin Zhao}

\affiliation[a]{Department of Modern Physics, and Anhui Center for fundamental sciences in theoretical physics, \\
University of Science and Technology of China, Hefei, Anhui 230026, China}
\affiliation[b]{ PRISMA$^+$ Cluster of Excellence and Mainz Institute for Theoretical Physics\\
Johannes Gutenberg University, 55099 Mainz, Germany}
\affiliation[c]{ McGill University Department of Physics and Trottier Space Institute,
3600 Rue University, Montreal, QC, H3A 2T8, Canada}
\affiliation[d]{Bethe Center for Theoretical Physics and Physikalisches Institut,
 Universität Bonn\\
Nussallee 12, 53115 Bonn, Germany}
\affiliation[e]{International Centre for Theoretical Physics Asia-Pacific, University of Chinese Academy of Sciences,
100190 Beĳing, China\footnote{Address after October 1, 2025.}}

\emailAdd{dinggj@ustc.edu.cn}
\emailAdd{siichiang@mail.ustc.edu.cn}
\emailAdd{yonxu@uni-mainz.de}
\emailAdd{wenbin.zhao@uni-bonn.de}

\abstract{We use modular symmetry as an organizing principle that attempts to simultaneously address the lepton flavor puzzle, inflation, and post-inflationary reheating. We demonstrate this approach using the finite modular group $A_4$ in the lepton sector. In our model, neutrino masses are generated via the Type-I see-saw mechanism, with modular symmetry dictating the form of the Yukawa couplings and right-handed neutrino masses. The modular field also drives inflation, providing an excellent fit to recent Cosmic Microwave Background (CMB) observations. The corresponding prediction for the tensor-to-scalar ratio is very small, $r \sim \mathcal{O}(10^{-7})$, while the prediction for the running of the spectral index, $\alpha \sim -\mathcal{O}(10^{-3})$, could be tested in the near future. An appealing feature of the setup is that the inflaton-matter interactions required for reheating naturally arise from the expansion of relevant modular forms. Although the corresponding inflaton decay rates are suppressed by the Planck scale, the reheating temperature can still be high enough to ensure successful Big Bang nucleosynthesis. The same couplings responsible for reheating can also contribute to generating baryon asymmetry of the Universe through non-thermal leptogenesis. However, the contribution is negligibly small in the current inflationary setup.}

\begin{document}
\begin{flushright}
    MITP-24-084
\end{flushright}
\maketitle
\flushbottom

\section{Introduction}
Inflation is an elegant paradigm that resolves the flatness, horizon, and monopole problems~\cite{Starobinsky:1980te, Guth:1980zm, Linde:1981mu, Albrecht:1982wi}. The simplest inflationary model is the so-called slow-roll  single-field inflation, where a scalar inflaton field $\phi $ gradually rolls down its potential~\cite{Martin:2013tda}. To match the results from  recent cosmic microwave background (CMB) experiments~\cite{Planck:2018jri, BICEP2:2018kqh}, the inflaton potential has to be sufficiently flat. Furthermore, according to the latest Planck data~\cite{Planck:2018jri}, the most favored single-field inflation models are those with concave potentials, where $V^{\prime \prime}(\phi)<0 $. One example is hilltop inflation \cite{Boubekeur:2005zm}. It is shown recently  that  hilltop-like inflation can be realized in the framework of modular symmetry with the modular field acting as the inflaton~\cite{Ding:2024neh,King:2024ssx}.  See Refs.~\cite{Schimmrigk:2014ica,Schimmrigk:2015qju,Schimmrigk:2016bde,Kobayashi:2016mzg,Abe:2023ylh,Frolovsky:2024xet_pb,Casas:2024jbw,Kallosh:2024ymt,Kallosh:2024pat, Aoki:2024ixq} for inflationary scenario based on modular symmetry\footnote{The hybrid inflation induced by right-handed sneutrino with modular flavor symmetry was discussed in~\cite{Gunji:2022xig}.}.

Due to the exponential expansion, the Universe at the end of inflation is non-thermal. Any viable inflationary scenario must also explain how the Universe reheats and achieves thermal equilibrium with a temperature above the MeV scale, necessary for Big Bang Nucleosynthesis (BBN)~\cite{Kawasaki:2000en, Hannestad:2004px, deSalas:2015glj, Hasegawa:2019jsa}. In this work, we revisit modular slow-roll hilltop inflation~\cite{Ding:2024neh} with a particular focus on the postinflationary reheating process. Notably, we find that inflaton-matter coupling required for reheating  naturally arises from the modular symmetry approach to solve the flavor puzzles~\cite{Feruglio:2017spp}. In the framework of using modular symmetry to explain lepton masses and mixing angles, it is required that the Yukawa couplings be modular forms which are holomorphic functions of the complex modulus $\tau$; see Refs.~\cite{Kobayashi:2023zzc,Ding:2023htn} for recent reviews. This naturally gives rise to couplings between the inflaton field and SM particles, which facilitate the production of these particles and reheat the Universe following inflation. Analyzing reheating after modular slow-roll inflation is one of the main objectives of this work.

We briefly outline our approach. Since the modular forms and Yukawa couplings are determined by the vacuum expectation value (VEV) of the modulus field, our primary objective is to construct a scalar potential that supports inflation and has a minimum at the required VEV, which also fits the lepton data. The process of dynamically fixing the VEV of the modulus field is referred to as modulus stabilization. It has been shown that the extrema of modular invariant scalar potentials are typically located near the boundary of the fundamental domain or along the imaginary axis~\cite{Cvetic:1991qm,Kobayashi:2019xvz,Kobayashi:2019uyt,Ishiguro:2020tmo,Novichkov:2022wvg,Leedom:2022zdm,Knapp-Perez:2023nty,King:2023snq,Kobayashi:2023spx,Higaki:2024jdk}. Flavor models with VEV around the fixed points $\tau=i,\, \tau=\omega=e^{i 2\pi/3}$ are particularly interesting to us, as a small deviation from the fixed point can be used to naturally explain the lepton mass hierarchy and CP violation~\cite{Feruglio:2021dte,Novichkov:2021evw,Feruglio:2022koo,Feruglio:2023mii,Ding:2024xhz}. Inspired by this, we investigate the possibility that the inflaton slowly rolls from $i$ and oscillates around a point near $\omega$.

To provide a concrete example, we construct a model with $A_4 $ modular symmetry. In this model, light neutrino masses are generated via the Type-I seesaw mechanism. Modular symmetry requires the mass terms for the right-handed neutrinos (RHNs) to be modular forms, which also induce couplings between the inflaton and RHNs. We find that inflaton dominantly decays into RHNs after inflation. Although the corresponding inflaton decay rates are suppressed by the Planck scale, the reheating temperature can still be high enough to ensure successful Big Bang nucleosynthesis (BBN). We also find that the reheating temperature is lower than the RHN mass scale, which gives the possibility to generate the baryon asymmetry in the universe via  non-thermal leptogenesis~\cite{Lazarides:1990huy, Asaka:1999yd, Asaka:1999jb, Senoguz:2003hc, Hahn-Woernle:2008tsk, Antusch:2010mv, Drees:2024hok}.

Our results suggest that modular symmetry can be a good organising principle to solve flavor puzzle, inflation as well as postinflationary reheating. Modular invariant models for flavor problems are hard to distinguish from each other since they give more or less the same predictions in lepton masses and mixing patterns. However, their cosmological implications might be different and offer another window to probe modular symmetry.

The article is organized  as follows. In section \ref{sec:Modular_Intro}, we offer a brief introduction for modular symmetry. A specific lepton flavor model $A_4$ is presented in section~\ref{sec:A4}.  We focus on modular slow-roll inflation and the post-inflationary reheating in section~\ref{sec:inflation}. Our main results are summarized in section \ref{sec:conclusion}. We give a short introduction to modular group $\Gamma_3$ in Appendix~\ref{app:A4-MF}. The vacuum structure of the scalar potential at the fixed point $\tau=\omega$ and global minimum $\tau=\tau_0$  are investigated in Appendix~\ref{app:modular_derivative}. The two-body and three-body decays of the inflaton are studied in Appendix~\ref{sec:3-body}. The baryon asymmetry generated through non-thermal leptogenesis is discussed in Appendix~\ref{app:leptogenesis}. Throughout this work, we use the notation $ M_{\text{Pl}} \equiv \frac{1}{\sqrt{8 \pi G}} = 2.4 \times 10^{18}~\text{GeV}$, with $G$  being the Newton constant, corresponding to the reduced Planck scale. 

\section{Modular symmetry and modular invariant theory }\label{sec:Modular_Intro}

The full modular group $\Gamma$ is the group of the linear fraction transformations acting on the complex modulus $\tau$ in the upper-half complex plane as follow,
\begin{equation}
\label{eq:linear-frav-trans}\tau\rightarrow\gamma\tau=\frac{a\tau+b}{c\tau+d},~~~~~\texttt{Im}(\tau)>0\,,
\end{equation}
where $a$, $b$, $c$ and $d$ are integers satisfying $ad-bc=1$. Hence, each linear
fractional transformation is associated with a two-dimensional integer matrix $\gamma=\begin{pmatrix}
a ~& b \\
c ~& d
\end{pmatrix}$
with unit determinant. Since $\gamma$ and $-\gamma$ lead to the same linear fraction transformation, the modular group $\overline{\Gamma}$ is isomorphic to the projective special linear group  $PSL(2, \mathbb{Z})\equiv SL(2, \mathbb{Z})/\left\{\pm \mathds{1}\right\}$. The modular group $\overline{\Gamma}$ can be generated by the duality transformation $\mathcal{S}: \tau\rightarrow-1/\tau$ and the shift transformation
$\mathcal{T}: \tau\rightarrow\tau+1$ which are represented by the following matrices
\begin{equation}\label{eq:ST}
\mathcal{S}=\begin{pmatrix}
0 ~& 1 \\
-1~& 0
\end{pmatrix},~~~\mathcal{T}=\begin{pmatrix}
1~& 1\\
0 ~& 1
\end{pmatrix}\,.
\end{equation}
The linear fraction transformations can map the upper half complex plane into the fundamental domain for which two interior points are not related with each other under eq.~\eqref{eq:linear-frav-trans}. The standard fundamental domain $\mathcal{D}$ denotes the set
\begin{equation}
\mathcal{D}=\left\{\tau\Big| \left|\tau\right|\geq1, \left|\texttt{Re}(\tau)\right|\leq1/2~ \text{and} ~\texttt{Im}(\tau)>0\right\}\,.
\end{equation} 
Under the action of $\gamma\in\Gamma$, the  chiral supermultiplets $\Phi_I$ of matter fields transform as
\begin{equation}
\Phi_I\xrightarrow{\gamma}(c\tau+d)^{-k_I}\rho_I(\gamma)\Phi_I\,,
\end{equation} 
where $k_I$ is the modular weight of $\Phi_I$, and $\rho_I$ is a unitary representation of the finite modular group
$\Gamma_N=SL(2, \mathbb{Z})/\pm\Gamma(N)$ or $\Gamma'_N=SL(2, \mathbb{Z})/\Gamma(N)$. $\Gamma(N)$ is a principal congruence subgroup of $SL(2, \mathbb{Z})$ and the level $N$ is fixed to be  certain positive integer. We work in the framework of $N=1$  supergravity including the K\"ahler modulus $\tau$, one dilaton $S$ and the matter fields $\Phi_I$ of the Standard Model\footnote{In this section we use the Planck units where the reduced Planck
mass $M_{\text{Pl}}=1$.}. The K\"{a}hler potential $\mathcal{K}$ and the superpotential $\mathcal{W}$ combine into an function $\mathcal{G}$,
\begin{equation}
\mathcal{G}(\tau, \bar{\tau};S, \bar{S}; \Phi_I, \bar{\Phi}_I)=\mathcal{K}(\tau, \bar{\tau};S, \bar{S}; \Phi_I, \bar{\Phi}_I)+\ln\left|\mathcal{W}(\tau, S,  \Phi_I )\right|^2\,.
\end{equation}
The modular transformations of $\mathcal{K}$ and $\mathcal{W}$ compensate with each other so that the function $\mathcal{G}$ is modular invariant. For a given K\"ahler potential $\mathcal{K}$ and superpotential $\mathcal{W}$, the relevant part of the interaction Lagrangian reads~\cite{Wess:1992cp}:
\begin{eqnarray}\label{eq:lagrangian}
\nonumber e^{-1}\mathcal{L} &=& -\mathcal{K}_{\alpha\overline{\beta}}( \partial_\mu \phi^\alpha \partial^\mu \Bar{\phi}^{\overline{\beta}}+ i \Bar{\chi}^{\overline{\beta}}\Bar{\sigma}^\mu \partial_\mu \chi^\alpha)-e^{{\cal K}}\left({\cal K}^{\alpha\overline{\beta}}D_{\alpha}{\cal W}\overline{D_{\beta}{\cal W}} - 3|{\cal W}|^2\right)\\
&&-\frac{1}{2}e^{{\cal K}/2}\left[ (D_\alpha D_\beta {\cal W}) \chi^\alpha \chi^\beta + \text{h.c.}\right]\,,
\end{eqnarray}
where we denote the scalar field as $\phi^\alpha$ and the 2 component spinor field as $\chi^\alpha$, where $\alpha$ runs over all the chiral supermultiples in the theory, including $\tau, S$ and $\Phi_I$. $e=\sqrt{-\det g}$ is the determinant of frame field, and $D_\alpha {\cal W} \equiv \partial_\alpha {\cal W} + {\cal W} (\partial_\alpha {\cal K})$ denotes the covariant derivative and ${\cal K}^{\alpha\overline{\beta}}$ is the inverse of the K\"ahler metric ${\cal K}_{\alpha\overline{\beta}}=\partial_\alpha\partial_{\overline{\beta}} {\cal K}$. 
We adopt the following form of the K\"ahler potential ${\cal K}$
\begin{eqnarray}
\mathcal{K}(\tau, \bar{\tau};S, \bar{S}; \Phi_I, \bar{\Phi}_I) &=& K(S, \bar{S})- 3\ln\left[-i(\tau-\bar{\tau})\right] +  \sum_I(-i \tau + i \bar{\tau})^{-k_I}|\Phi_I|^2\,,
\end{eqnarray}
where we take the Planck units with the reduced Planck
mass $M_{\text{Pl}}=1$, and the K\"ahler potential for the dilaton 
\begin{equation}
K(S,\bar{S})=-\ln(S+\bar{S})+\delta k_{\text{np}}(S,\bar{S})\,.
\end{equation}
Here, $\delta k_{\text{np}}$ denotes the additional corrections from some stringy effects such as Shenker-like effects~\cite{Shenker:1990uf}, and it is necessary for the dilaton stabilization.  Since  $(-i\tau+i\bar{\tau})\xrightarrow{\gamma} |c \tau + d|^{-2}(-i\tau+i\bar{\tau})$, the modular transformation of the  K\"ahler potential ${\cal K}$ is
\begin{eqnarray}
\mathcal{K}\xrightarrow{\gamma}\mathcal{K} + 3\ln(c \tau + d) + 3\ln(c \bar{\tau} + d)\,.
\end{eqnarray}
The superpotential $\mathcal{W}$ is a holomorphic function of $\tau$, $S$ and $\Phi_I$, and it can be written as
\begin{eqnarray}
\mathcal{W}(\tau, S,  \Phi_I ) &=& \mathcal{W}_{\text{moduli}}(\tau,S) + \mathcal{W}_{\text{matter}}(\tau, \Phi_I)\,.
\end{eqnarray}
It is required that  $\mathcal{W}$ has to be a modular function of weight $-3$, i.e.,
\begin{equation}\label{eq:suptrans}
{\cal W} \xrightarrow{\gamma} e^{i\delta(\gamma)}(c \tau + d)^{-3}{\cal W} \,,
\end{equation}
where the phase $\delta(\gamma)$ depending on the modular transformation $\gamma\in PSL(2, \mathbb{Z})$ is the so-called multiplier system.

As shown in~\cite{Cvetic:1991qm}, the superpotential $\mathcal{W}_{\text{moduli}}$ can in general
be parameterized as,
\begin{equation}\label{eq:superpotential}
\mathcal{W}_{\text{moduli}}(S,\tau)=\Lambda^3_W\frac{\Omega(S)H(\tau)}{\eta^6(\tau)}\,,
\end{equation}
where $\Lambda_W$ is an energy scale, $\eta(\tau)$ is the Dedekind $\eta$ function given in eq.~\eqref{eq:eta-func}, and $\Omega(S)$ is an arbitrary function of the $S$-field. Here we assume that the dilaton field $S$ is stabilized. The modular function $H(\tau)$ is regular in the fundamental domain, and it has the following parameterization~\cite{Cvetic:1991qm}:

\begin{equation}
H(\tau)=(j(\tau)-1728)^{m/2}j(\tau)^{n/3}\mathcal{P}(j(\tau))\,,
\label{eq:H1new}
\end{equation}
where $j(\tau)$ denotes the modular invariant $j$ function, $\mathcal{P}(j(\tau))$ is  an arbitrary polynomial function of $j(\tau)$, and both $m$ and $n$ are non-negative integers.

The matter superpotential $\mathcal{W}_{\text{matter}}$ can be expanded in power series of the supermultiplets $\Phi_I$ as follow,
\begin{equation}
\mathcal{W}_{\text{matter}}(\tau, \Phi_I)=\sum_n Y_{I_1\ldots I_n}(\tau) \Phi_{I_1}\ldots  \Phi_{I_n}\,,
\end{equation}
where $Y_{I_1\ldots I_n}(\tau)$ is a modular form multiplet and it should transform as,
\begin{equation}
Y_{I_1\ldots I_n}(\tau)\xrightarrow{\gamma} Y_{I_1\ldots I_n}(\gamma\tau)= e^{i\delta(\gamma)}(c \tau + d)^{ k_{Y}} \rho_Y(\gamma) Y_{I_1\ldots I_n}(\tau)\,,
\end{equation}
with
\begin{equation}
\begin{cases}
k_Y=k_{I_1} + \ldots + k_{I_n} - 3\,,\\
\rho_Y \bigotimes \rho_{I_1}\bigotimes\ldots \bigotimes  \rho_{I_n} \ni \textbf{1}\,,
\end{cases}
\end{equation}
where $\mathbf{1}$ refers to the trivial singlet of $\Gamma_N$ or $\Gamma'_N$.

%%%%%%%%%%%%%%%%%%%%%%%%%%%%%%%%%%%%%%%%%%%%%%%%%%%%%%%%%%%%%%%%%%%%%%
\section{Lepton flavor model with $\Gamma_3\cong A_4$ symmetry} \label{sec:A4}
%%%%%%%%%%%%%%%%%%%%%%%%%%%%%%%%%%%%%%%%%%%%%%%%%%%%%%%%%%%%%%%%%%%%%%
In the following, we focus primarily on a specific model with $ A_4$ symmetry, and the light neutrino masses are generated by the Type-I seesaw mechanism. We will present the lepton sector and omit the quark sector, since the modulus $\tau$ has the largest couplings with the right-handed neutrinos because of their heavy masses. The quark sector contributes sub-dominantly to reheating process.  The model is specified by the following representation assignments and modular weights of the lepton fields:
\begin{eqnarray}\nonumber
&&L\sim \mathbf{3}\,,~~e^{c}\sim \mathbf{1'}\,,~~\mu^{c}\sim \mathbf{1'}\,,~~\tau^{c}\sim \mathbf{1''}\,,~~N^{c}\equiv \{N^{c}_{1},N^{c}_{2},N^{c}_{3}\}\sim \mathbf{3}\,,\\
&&k_{L}=1\,,~~k_{e^{c}}=1\,,~~k_{\mu^{c}}=5\,,~~k_{\tau^{c}}=5\,,~~k_{N}=1\,.
\end{eqnarray}
The two Higgs superfields $H_u$ and $H_d$ transform trivially under modular symmetry. The modular invariant superpotentials responsible for the mass of lepton are
\begin{equation}
\mathcal{W}_{\text{matter}} = \mathcal{W}_{E} + \mathcal{W}_{\nu}\,,
\end{equation}
where\footnote{In this work, the superpotential $\mathcal{W}_{\text{matter}}$ should have the same modular transformation property as $\mathcal{W}_{\text{moduli}}$. Thus unlike the global SUSY scenario, we introduce an extra function $\eta^{6}$ in the matter superpotential.}
\begin{small}
\begin{eqnarray}
\nonumber
\eta^6\mathcal{W}_{E}&=&y_{1} e^{c}(LY^{(2)}_{\mathbf{3}})_{\bm{1''}}H_{d}
+ y_{2} \mu^{c}(LY^{(6)}_{\bm{3}I})_{\bm{1''}}H_{d}
+ y_{3} \mu^{c}(LY^{(6)}_{\bm{3}II})_{\bm{1''}}H_{d}\\
\nonumber &&+ y_{4} \tau^{c}(LY^{(6)}_{\bm{3}I})_{\bm{1'}}H_{d}+y_{5}  \tau^{c}(LY^{(6)}_{\bm{3}II})_{\bm{1'}}H_{d}\,,\\  
\eta^6\mathcal{W}_{\nu}&=&g_1\left((N^{c}L)_{\mathbf{3}_{S}}Y^{(2)}_{\mathbf{3}}\right)_{\mathbf{1}}H_{u}+g_2\left((N^{c}L)_{\mathbf{3}_{A}}Y^{(2)}_{\mathbf{3}}\right)_{\mathbf{1}}H_{u}+\frac{1}{2} \Lambda_N \left( (N^{c}N^{c})_{\mathbf{3}_{S}}Y^{(2)}_{\mathbf{3}}\right)_{\mathbf{1}}\,.
\end{eqnarray}
\end{small}
Here, the couplings $y_1$, $y_2$, $y_4$ and $\Lambda_N$ can be taken to be real since their phases can be absorbed by field redefinition, while the phases of $y_3$, $y_5$ and $g_2$ can not be removed. The definitions of modular forms $Y^{(2)}_{\mathbf{3}},Y^{(6)}_{\bm{3}I},Y^{(6)}_{\bm{3}II}$ and group contractions can be found in Appendix~\ref{app:A4-MF}. The Yukawa term for leptons and neutrinos, expressed in the flavor basis, can be written as
\begin{eqnarray}
{\cal L} =  {\cal Y}_{E}^{ij}(\tau)L^c_i L_j H_d + {\cal Y}_{D}^{ij}(\tau)N^c_i L_j H_u + \frac{1}{2} {\cal Y}_{N}^{ij}(\tau) N^c_i N^c_j  + \text{h.c.}\,,
\end{eqnarray} 
where the ${\cal Y}_{E}^{ij}$, ${\cal Y}_{D}^{ij}$ and ${\cal Y}_{N}^{ij}$ are some linear combinations of modular forms, and $i,j=1, 2, 3$ are indices of generation. The corresponding charged lepton and neutrino mass matrices read as
%%%%%%%%%%%%%%%%%%%%%%%%%%%%%%%%%%%%%
\begin{small}
\begin{eqnarray} \nonumber
  M_{E}&=&v_d {\cal Y}_{E}(\tau)=\left(
\begin{array}{ccc}
 y_{1}Y_{\bm{3},3}^{(2)} ~&~ y_{1}Y_{\bm{3},2}^{(2)} ~&~ y_{1}Y_{\bm{3},1}^{(2)} \\
 y_{2} Y_{\bm{3}I,3}^{(6)}+y_{3} Y_{\bm{3}II,3}^{(6)} ~&~ y_{2} Y_{\bm{3}I,2}^{(6)}+y_{3} Y_{\bm{3}II,2}^{(6)} ~&~ y_{2} Y_{\bm{3}I,1}^{(6)}+y_{3} Y_{\bm{3}II,1}^{(6)} \\
 y_{4} Y_{\bm{3}I,2}^{(6)}+y_{5} Y_{\bm{3}II,2}^{(6)} ~&~ y_{4} Y_{\bm{3}I,1}^{(6)}+y_{5} Y_{\bm{3}II,1}^{(6)} ~&~ y_{4} Y_{\bm{3}I,3}^{(6)}+y_{5} Y_{\bm{3}II,3}^{(6)} \\
\end{array}
\right)\frac{v_{d}\eta_0^6}{\eta^6}\,,\\
\nonumber  M_{D} &=&v_u {\cal Y}_{D}(\tau) = \left(
\begin{array}{ccc}
 2 g_1Y_{\bm{3},1}^{(2)}  ~&~- g_1Y_{\bm{3},3}^{(2)}-g_2 Y_{\bm{3},3}^{(2)}   ~&~  - g_1Y_{\bm{3},2}^{(2)}+g_2 Y_{\bm{3},2}^{(2)}   \\
- g_1Y_{\bm{3},3}^{(2)} +g_2 Y_{\bm{3},3}^{(2)} ~&~ 2 g_1Y_{\bm{3},2}^{(2)}  ~&~ - g_1Y_{\bm{3},1}^{(2)} - g_2 Y_{\bm{3},1}^{(2)} \\
- g_1Y_{\bm{3},2}^{(2)} - g_2 Y_{\bm{3},2}^{(2)}   ~&~   - g_1Y_{\bm{3},1}^{(2)}+g_2 Y_{\bm{3},1}^{(2)}  ~&~ 2 g_1Y_{\bm{3},3}^{(2)}  \\
\end{array}
\right)\frac{v_{u}\eta_0^6}{\eta^6} \,,\\
 M_{N}&=&  \Lambda_N   {\cal Y}_{N}(\tau) = \left(
\begin{array}{ccc}
2 Y_{\bm{3},1}^{(2)} ~&~ - Y_{\bm{3},3}^{(2)}  ~&~ - Y_{\bm{3},2}^{(2)} \\
 - Y_{\bm{3},3}^{(2)}  ~&~ 2 Y_{\bm{3},2}^{(2)} ~&~ - Y_{\bm{3},1}^{(2)} \\
 - Y_{\bm{3},2}^{(2)}  ~&~ - Y_{\bm{3},1}^{(2)}  ~&~ 2 Y_{\bm{3},3}^{(2)}  \\
\end{array}
\right)\frac{\Lambda_N \eta_0^6}{\eta^6}\,,
\end{eqnarray}
\end{small}
where we have rescaled parameters $y_1,g_1,\Lambda_N$ by $\eta_0^6$ to compensate the existence of $\eta^6$, and $\eta_0$ is defined at the benchmark point $\eta_0=\eta(\tau_0)$. Fitting results will be the same as the global SUSY case. The light Majorana neutrino mass matrix is obtained through the Type-I seesaw as follows:
\begin{equation}
m_{\nu}=-M^{T}_DM^{-1}_NM_D\,.
\end{equation}
We can perform the transformation from the flavor basis to the mass basis by
\begin{eqnarray}
\nonumber U^{e\,T}_R M_E U^e_L &=&\text{diag}(m_e,m_\mu,m_\tau)\,,\\
U^{\nu\,T}_L m_{\nu} U^{\nu}_L &=&\text{diag}(m_{1},m_{2},m_{3})\,,\\
\nonumber U^{T}_N M_N U_N &=&\text{diag}(M_{1},M_{2},M_{3})\,,
\end{eqnarray}
where $U^{e}_L$, $U^{e}_R$, $U^{\nu}_L$ and $U_N$ are unitary matrices; $m_i$ and $M_i$ denote the masses for active neutrinos and right handed neutrinos, respectively. We consider $\langle\tau\rangle$ to stay at the arc, i.e. $\abs{\tau}=1$. We use the following benchmark values of the free parameters:
\begin{equation}\label{eq:parameters}
\begin{gathered}
\quad \tau_0=\langle\tau\rangle=- 0.4847 +  0.8747 i\,,\\
y_{2}/y_{1}=5.1844\times 10^{2} \,,\quad y_{3}/y_{1}=1.46 59e^{-2.4143i}\times 10^{2}\,, \\
 \quad y_{4}/y_{1}=2.3952\times 10^{4}\,,\quad
y_{5}/y_{1}=1.2117e^{-0.5024i}\times 10^{3}\,,\\
g_2/g_1=0.2465 \,, \quad
y_{1}v_{d} =0.2499 ~\text{MeV}\,,\quad \frac{(g_1v_{u})^{2}}{\Lambda_N}= 19.9673~\text{meV}\,.
\end{gathered}
\end{equation}
Notice that $\tau_0$ is close to the modular symmetry fixed point $\omega\equiv e^{2\pi i/3}$.
The corresponding observables for mixing parameters of leptons and masses are given by
\begin{eqnarray}
\nonumber&& \sin^{2}\theta_{12}=0.307\,,\quad \sin^{2}\theta_{13}=0.022\,,\quad \sin^{2}\theta_{23}=0.454\,,\\
\nonumber&&  \delta_{CP}/\pi=0.855\,,\quad \alpha_{21}/\pi=0.939\,,\quad \alpha_{31}/\pi=0.271\,,\\
\nonumber&&  m_e/m_{\mu}=0.00474\,,\quad m_{\mu}/m_{\tau}=0.0588\,, \quad \frac{\Delta m_{21}^{2}}{\Delta m_{31}^{2}} = 0.0296\,,\\ \nonumber&& m_1=25.725~\text{meV}\,,\quad m_2=27.127~\text{meV}\,,\quad m_3=56.274~\text{meV}\,,\\ && m_{\beta\beta}=9.615 ~\text{meV}\,,\quad (M_1,M_2,M_3)=\Lambda_N(1.372,1.447,2.818)\,,
\end{eqnarray}
where $m_{\beta\beta}$ is the effective mass in neutrinoless double beta decay, $M_{1,2,3}$ are the masses of heavy right-handed neutrinos. All the above lepton masses and mixing angles are within $1\sigma$ region of the experimental data~\cite{ParticleDataGroup:2024cfk,Esteban:2024eli}. Remarkably, $M_1$ and $M_2$ are quasi-degenerate, which plays a crucial role for leptogenesis.

\section{Inflationary cosmology of modular invariance}\label{sec:inflation}
In this section, we briefly review inflation \cite{Baumann:2009ds} and its predictions for Cosmic Microwave Background (CMB) observables. The simplest scenario is slow-roll inflation, in which the inflaton field slowly rolls down the potential \( V(\phi) \). To connect with observables, we define the slow-roll parameters:
\begin{align}\label{eq:SR_parameters}\epsilon_V = \frac{M_{\text{Pl}}^2}{2}\left(\frac{V'}{V}\right)^2 \,,~~
\eta_V = M_{\text{Pl}}^2\left(\frac{V''}{V}\right)\,,~~
\xi^2_V = M_{\text{Pl}}^4\left(\frac{V'V'''}{V^2}\right)\,,
\end{align}
where $\prime$ denotes the derivative of the potential with respect to $\phi$. For slow-roll  inflation, it is required that $ \epsilon_V$, $|\eta_V|$ and $|\xi_V^2| \ll 1$ during inflation.  
The end of inflation is defined to be at a field value $\phi_{\text{end}}$ such that $\text{max}\left\{\epsilon_V(\phi_{\text{end}}),|\eta_V(\phi_{\text{end}})|\,, |\xi_V^2(\phi_{\text{end}})|\right\} =1$.
Moreover, to effectively resolve the flatness problem, the exponential expansion must last sufficiently long, which can be quantified by the number of e-folds $N_e$.  Under the slow-roll approximation, it can be expressed as
\begin{align}
N_e(\phi_*)=\int_{\phi_*}^{\phi_{\text{end}}}\frac{1}{\sqrt{2\epsilon_V}} \frac{d \phi}{M_{\text{Pl}}}\,,
\end{align}
where $\phi_*$ denotes the field value when the CMB pivot scale
$k_{\star} = 0.05 \rm{Mpc}^{-1}$ first crossed out the horizon.
The predictions for the power spectrum of the curvature perturbation $\mathcal{P_R}$, spectral index $n_s$, its running $\alpha\equiv \frac{d n_s}{d \log k}$, and the tensor-to-scalar ratio $r$ during slow-roll inflation are given by \cite{Lyth:2009zz}:
\begin{align}
\mathcal{P_R} = \frac{V}{24\pi^2 \epsilon_V}\,,~~
  n_s = 1 - 6 \epsilon_V + 2\eta_V\,,~~
  \alpha = 16 \epsilon_V\eta_V - 24  \epsilon_V^2 - 2\xi_V^2\,,~~
   r = 16  \epsilon_V\,,
\end{align}
respectively.
The recent Planck 2018 measurements plus results on baryonic acoustic oscillations (BAO)  at the pivot scale
$k_{\star} = 0.05 \rm{Mpc}^{-1}$, give \cite{Planck:2018vyg}:
\begin{equation}  \label{planck2018}
\mathcal{P_R} = (2.1 \pm 0.1)\times 10^{-9}\,,~   n_s =  0.9659  \pm 0.0040\,,~ \alpha = -0.0041 \pm 0.0067\,.
\end{equation} 
For the tensor-to-scalar ratio $r$, BICEP/Keck 2018 offers  most stringent bound \cite{BICEP:2021xfz}, which is
\begin{equation}  \label{BK2018}
r < 0.036\,\quad \text{at 95\% \text{C.L.}}\,.
\end{equation} 
The next-generation CMB experiments for example CORE~\cite{COrE:2011bfs}, AliCPT~\cite{Li:2017drr}, LiteBIR~ \cite{Matsumura:2013aja,LiteBIRD:2025trg}, CMB-S4~\cite{Abazajian:2019eic}  could reach  sensitivity of $r \sim \mathcal{O}(10^{-3})$.  
We note that the current constraint on the running $\alpha$ features a larger uncertainty. Future CMB measurements, such as those from CMB-S4, combined with investigations of smaller-scale structures—particularly through the Lyman-alpha forest—can refine constraints on the running of the spectral index to approximately $\alpha \sim \mathcal{O}(10^{-3})$ \cite{Munoz:2016owz}.

%%%%%%%%%%%%%%%%%%%%%%%%%%%%%%%%%%%%%%%%%%%%%%%%%%%%%%%%%%%%%%
\subsection{Modular invariant inflation}\label{sec:modular_inflation}
%%%%%%%%%%%%%%%%%%%%%%%%%%%%%%%%%%%%%%%%%%%%%%%%%%%%%%%%%%%%%%

In this section, we briefly discuss the modular invariant inflation model. This scenario has been studied recently in~\cite{Ding:2024neh, King:2024ssx}, where the inflationary trajectory follows the lower boundary of the fundamental domain between the two fixed points, $\tau = i$ and $\tau=\omega=e^{i2\pi/3}$. In this setup, modular symmetry plays a crucial role in ensuring the flatness of the inflationary potential and justifying the single-field approximation.

Although the fixed point $\omega$ is a promising candidate for the potential vacuum, the residual symmetry preserved at this point complicates addressing the lepton flavor problem within this framework. It has been noted that a slight deviation from this fixed point can naturally account for the lepton mass hierarchy and CP violation~\cite{Feruglio:2021dte,Novichkov:2021evw,Feruglio:2022koo,Feruglio:2023mii,Ding:2024xhz}. 

To embed inflation within the framework of modular invariance, we construct an inflationary potential with a minimum located at $\tau=\tau_0$ (cf. eq.~\eqref{eq:parameters}). Inflation occurs around the fixed point $\tau=i$ and then oscillates around the minimum of $\tau=\tau_0$  after inflation, during the reheating process. Building on the previous work~\cite{Ding:2024neh}, we continue to analyse the most general superpotential in eq.~\eqref{eq:superpotential}. Using eq.~\eqref{eq:lagrangian}, the scalar potential reads:
\begin{equation}
    V(\tau, S) = \Lambda^4 Z(\tau,\Bar{\tau}) \left[(A(S,\Bar{S})-3)\abs{H(\tau)}^2+\widehat{V}(\tau,\Bar{\tau})\right]\,.
\label{eq:simplifiedscalarpotential}
\end{equation}
Here, we define $\Lambda=(\Lambda^6_W  e^{K(S,\bar{S})}\abs{\Omega(S)}^2/M_{\text{Pl}}^2)^{1/4}$, representing the energy scale of the potential, which can be determined by CMB observables. The remaining terms are given by:
\begin{align}\label{AVZ}
\nonumber Z(\tau,\bar{\tau})&=\frac{1}{i(\tau-\bar{\tau})^3\abs{\eta(\tau)}^{12}}\,,\\
\nonumber A(S,\Bar{S})&=\frac{K^{S\bar{S}}D_S W D_{\Bar{S}}\bar{W}}{\abs{W}^2}=\frac{K^{S\bar{S}}\abs{\Omega_S+K_S\Omega}^2}{\abs{\Omega}^2}\,,\\
\widehat{V}(\tau,\bar{\tau})&=\frac{-(\tau-\bar{\tau})^2}{3}\abs{H_\tau(\tau)-\frac{3i}{2\pi}H(\tau)\widehat{G}_2(\tau,\bar{\tau})}^2\,.
\end{align}
where $H_\tau =\frac{\partial H}{\partial\tau}$ and the modified weight $2$ Eisenstein series $\widehat{G}_2(\tau)$ reads
\begin{equation}\label{eq:G2hat}
\widehat{G}_2(\tau)=G_2(\tau) + \frac{2\pi}{i(\tau-\bar{\tau})}\,, \quad G_2(\tau)= -4\pi i\frac{\partial_\tau\eta(\tau)}{\eta(\tau)}\,.
\end{equation}
We treat $A(S,\Bar{S})$ as a free parameter and use a special form of $H(\tau)$ to realize inflation:
\begin{equation}\label{eq:Hfun}
    H(\tau) = (j(\tau)-j(\tau_0))^2 \left[1+\beta\left(1-\frac{j(\tau)}{1728}\right)+\gamma\left(1-\frac{j(\tau)}{1728}\right)^2\right]\,,
\end{equation}
where the first part $(j(\tau)-j(\tau_0))^2$ is used to determine the vacuum position of the potential. In this setup, the scalar potential vanishes at $\tau=\tau_0$, as both $H(\tau_0)=H_{\tau}(\tau_0)=0$. Since we can ensure the potential is non-negative by setting $A(S,\Bar{S})>3$, $\tau_0$ is a Minkowski minimum of the potential. The rest ensures the flatness of the potential during inflation (around $\tau=i$). As we demonstrated in Appendix~\ref{app:modular_derivative}, $\tau=\omega$ becomes a local minimum, while $\tau=\tau_0$ is the global minimum of the potential.

In this paper, we adopt the inflationary trajectory along the arc~\cite{Ding:2024neh,King:2024ssx}. The modular field $\tau$ can be separated into radial and angular components, $\tau=\rho e^{i \theta}$. The corresponding kinetic term reads:
\begin{equation}
\mathcal{L}_{\text{kin}}=  -\frac{\partial^2\mathcal{K}}{\partial \tau \partial \bar{\tau}} \partial_{\mu} \tau \partial^{\mu} \bar{\tau}= 3\frac{M_{\text{Pl}}^2}{(-i\tau+i\bar{\tau})^2}\partial_{\mu} \tau \partial^{\mu} \bar{\tau}=\frac{3 M_{\text{Pl}}^2}{4 \sin^2(\theta)}\left( \frac{1}{\rho^2}\partial_{\mu} \rho \partial^{\mu} \rho+\partial_{\mu} \theta \partial^{\mu} \theta \right)\,,
\end{equation}
The modular invariance indicates the scalar potential in eq.~\eqref{eq:simplifiedscalarpotential} satisfies $\partial V /\partial \rho|_{\rho=1}=0$,which means $\dot{\rho}=0$ in the classical equation of motion along the arc. We have numerically checked that the effective mass term for $\rho$ field is much higher than the Hubble scale during inflation, suppressing its quantum excitation. Hereafter we will always set $\rho=1$ and $\Dot{\rho}=0$. These conditions decouple the angular and radial components. We will keep $\theta$ as the only dynamical degree of freedom during inflation. To normalize the kinetic term of $\theta$, we further introduce the canonical field $\phi$ 
\begin{equation}
\phi\equiv\sqrt{3/2}\, M_{\text{Pl}}\ln\left[\tan(\theta/2)\right]\,,
\end{equation}
which shall be understood as the canonical inflaton field giving rise to slow-roll inflation, whose minimum locates at $\phi_0$. We refer the reader to Refs.~\cite{Ding:2024neh,King:2024ssx} for more detailed analysis. Note one has to make $\tau=i$ a saddle point of potential to have single field inflation. This implies that our inflationary scenario is similar to the original hilltop inflation \cite{Boubekeur:2005zm}. 

For the convenience to obtain the inflationary predictions, we expand the full potential in eq.~\eqref{eq:simplifiedscalarpotential} around $\phi=0$ (corresponding to  $\tau=i$), leading to
\begin{equation}\label{eq:inflaton_potential}
 \quad V(\phi)=V_0\left(1-\sum_{n=1}^{\infty}C_{2n}\phi^{2n}\right)\,.
\end{equation}
The whole potential contains  $A(S,\Bar{S})$, $H(\tau)$ and $\hat{V}(\tau,\Bar{\tau})$, as shown in eq.~\eqref{eq:simplifiedscalarpotential}. We have assumed $A(S,\Bar{S})$ to be a constant, which serves as a free parameter in our model. The $H(\tau)$ function depends on the parameters $\beta,\gamma$ and $\hat{V}(\tau,\bar{\tau})$ depends on the $H(\tau)$ function and its derivative.

In terms of the expansion, all the coefficients $C_{2n}$ are determined by the parameters $A(S,\Bar{S})\,, \beta\,,\gamma$ appearing in the original potential shown in eq.~\eqref{eq:simplifiedscalarpotential}. Along the arc $\rho=1$, the $\mathcal{S}$ symmetry $\tau \to -1/\tau$, which is defined in eq.~\eqref{eq:ST}, indicates a
$Z_2$ symmetry in terms of the canonically normalised field $\phi \to -\phi$. We apply the slow-roll formalism as presented in the beginning of this section.

\begin{figure}
\centering
\begin{subfigure}[b]{1.\textwidth}
\centering
\includegraphics[width=\textwidth]{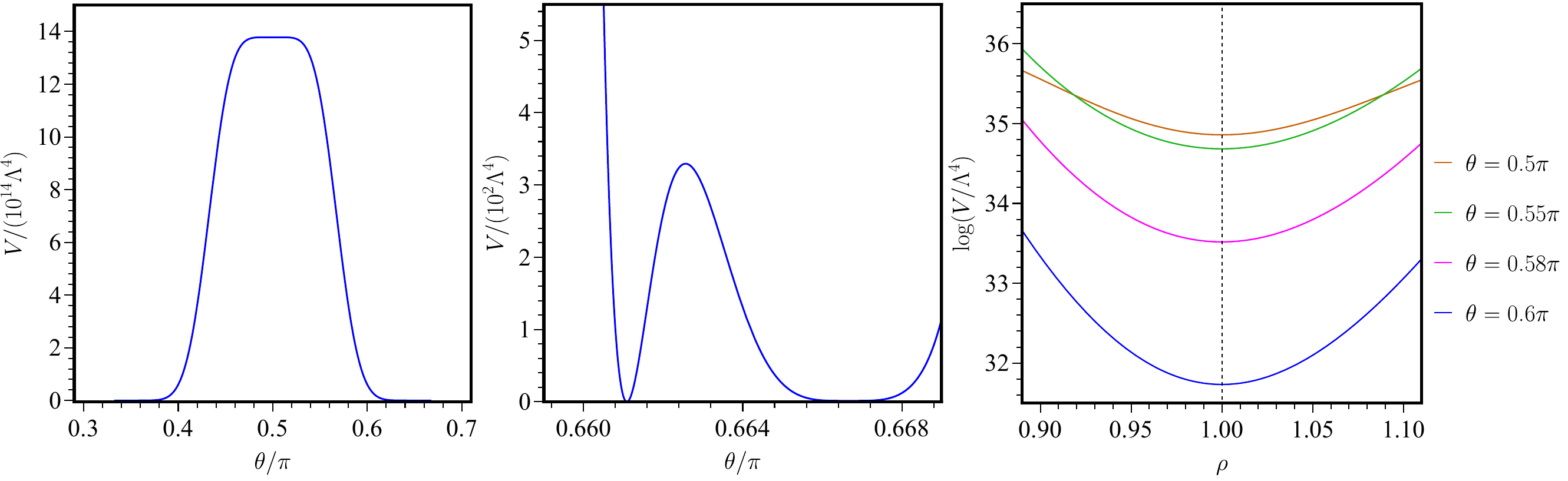}
\end{subfigure}
\hfill
\begin{subfigure}[b]{1.\textwidth}
\centering
\includegraphics[width=\textwidth]{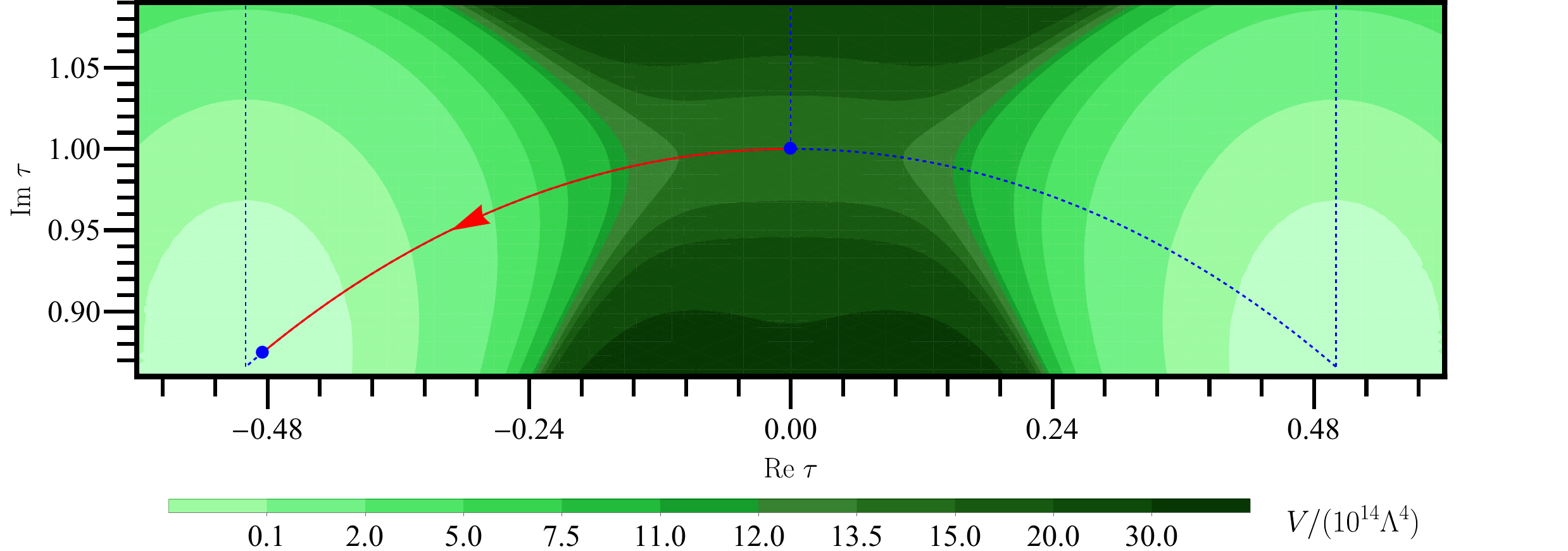}
\end{subfigure}
\caption{Shape of the potential along the angular and radial directions with $A = 55.2783$, $\beta = 0.6516$ and $\gamma=0$.  The top-left panel depicts the inflation potential with $\rho = 1$, where $\theta = \pi/2$ marks the starting point of inflation. The top-middle panel provides a zoomed-in view of the inflation potential around the desired minimum at $\tau = \tau_0$. Note that $\theta = 2\pi/3$ corresponds to a local minimum, whose potential energy does not vanish, while $\theta \approx 0.661\pi$ represents the global minimum. The top-right panel shows the radial potential with a fixed angular coordinate, where the inflationary trajectory remains at the minimum in this direction. Finally, the bottom panel is a contour plot of the inflation potential, with the red arrow indicating the trajectory of inflation.}
\label{fig:crosssecv2}
\end{figure}

\begin{figure}[t!]
\def\sepf{0.8}
\centering
\includegraphics[scale=\sepf]{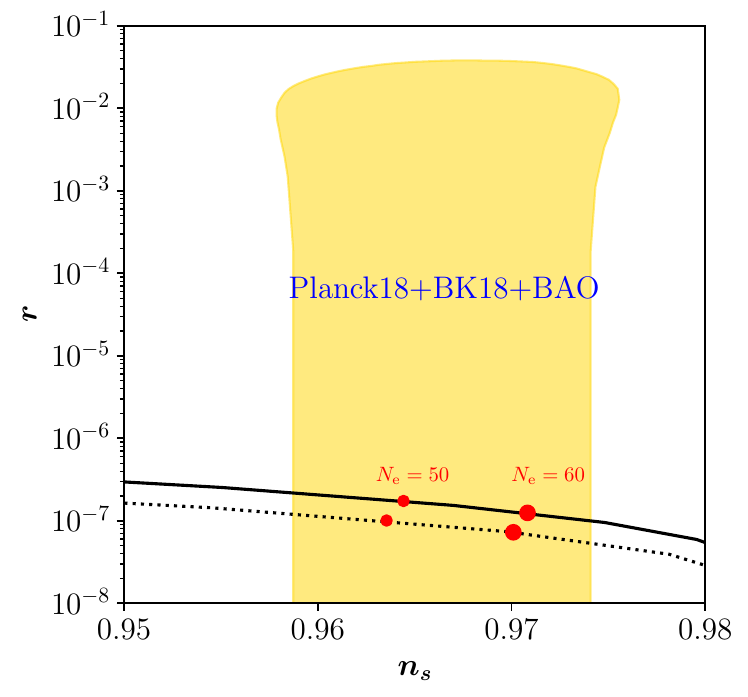}
\caption{The black lines represent the predictions for $(n_s, r)$ with model parameters:  $A =  55.2783$, $\beta = 0.6516$, $\gamma = 0$ (solid line) and $A =  80.2435$, $\beta = 0$, $\gamma = -1.2314$ (dotted line). The yellow shaded region corresponds to constraints from the combined results of Planck 2018, BICEP/Keck 2018, and BAO data \cite{BICEP:2021xfz}. 
The small and large red dots indicate $N_e = 50$ and $N_e = 60$, respectively.} 
\label{fig:ns-r}
\end{figure}

To show some representative examples for the inflationary predictions, we consider two benchmark parameters  below. The first one is the minimal case with model parameters
% %
\begin{equation}
\text{BP1:} \quad  A = 55.2783\,,  \quad \beta = 0.6516\,, \quad \gamma =0\,.  
\end{equation}
We show the shape of this potential along the radial and angular direction in Fig.~\ref{fig:crosssecv2}. Note $\tau=i$ is a saddle point of the potential. It is a local maximum in $\theta$ direction and a minimum in $\rho$ direction. Our inflation trajectory lies in the valley of radial direction. The prediction for $(n_s, r)$ is depicted by the black solid line in Fig.~\ref{fig:ns-r}, along with constraints from Planck 2018, BICEP/Keck 2018, and BAO data \cite{BICEP:2021xfz}. The two red dots correspond to $N_e = 50$ and $N_e = 60$, respectively. The predicted value of $r$ is of order $\mathcal{O}(10^{-7})$, a typical feature for small-field inflation models \cite{Martin:2013tda, Drees:2021wgd}. The prediction for $n_s$ lies within the $2\sigma$ region of the Planck 2018 results. Note that a larger $N_e$ implies that the inflaton field is closer to the saddle point, where the potential is flatter, resulting in a smaller $r$ and a more scale-invariant spectrum with a larger $n_s$. Consequently, $r$ decreases with increasing $n_s$ as $N_e$ increases, as shown by the black lines in Fig.~\ref{fig:ns-r}. The second benchmark example corresponds to
\begin{equation}
\text{BP2:} \quad  A = 80.2435\,, \quad \beta =0\,, \quad \gamma =-1.2314\,.
\end{equation}
The corresponding prediction is shown by the  black dotted line in Fig.~\ref{fig:ns-r} with a slightly smaller $r$. In both cases, we find $\alpha$ is of order $-\mathcal{O}(10^{-3})$, and the predictions for $n_s$, as shown in Fig.~\ref{fig:ns-r}, can lie within the $2\sigma$ range of the Planck 2018 results presented in eq.~\eqref{planck2018}. To illustrate the difference, we consider the central value of the spectral index $n_s =0.9659$. This corresponds to
\begin{equation}
r \approx 1.6\times10^{-7}\,, \quad \alpha \approx -7.078\times 10^{-4}\,,
\end{equation}
for BP1, and  
\begin{equation}
r \approx 9.0 \times10^{-8}\,, \quad \alpha \approx -7.083 \times 10^{-4}\,,
\end{equation}
for BP2. 
For the number of e-folds, we find that $N_e \simeq 52$ and $N_e \simeq 53$ for BP1 and BP2, respectively.   Note that both BP1 and BP2 correspond to $C_2 = 0.004$ and $C_4 =0$. The difference for the inflationary prediction for $r$ and $\alpha$ arises from higher order terms in the inflaton potential eq.~\eqref{eq:inflaton_potential}. 
Note that the two sets of benchmark parameters under consideration are representative of those that fit the CMB observables. For additional examples, we refer to \cite{Ding:2024neh}.

The prediction for $r$ is far below the sensitivity of next-generation CMB experiments, such as CMB-S4, which has a sensitivity of $r \sim \mathcal{O}(10^{-3})$ \cite{Abazajian:2019eic}. However, the prediction for a negative running $\alpha \sim -\mathcal{O}(10^{-3})$ could be tested within the sensitivity range of future CMB measurements, especially when combined with significantly improved investigations of structures at smaller scales, in particular the so-called Lyman-$\alpha$ forest \cite{Munoz:2016owz}.

%%%%%%%%%%%%%%%%%%%%%%%%%%%%%%%%%%%%%%%%%%%%%%%%%%%%%%%%%%%%%%%%%%%%%%%

\begin{figure}
\centering
\includegraphics[width=1.02\textwidth]{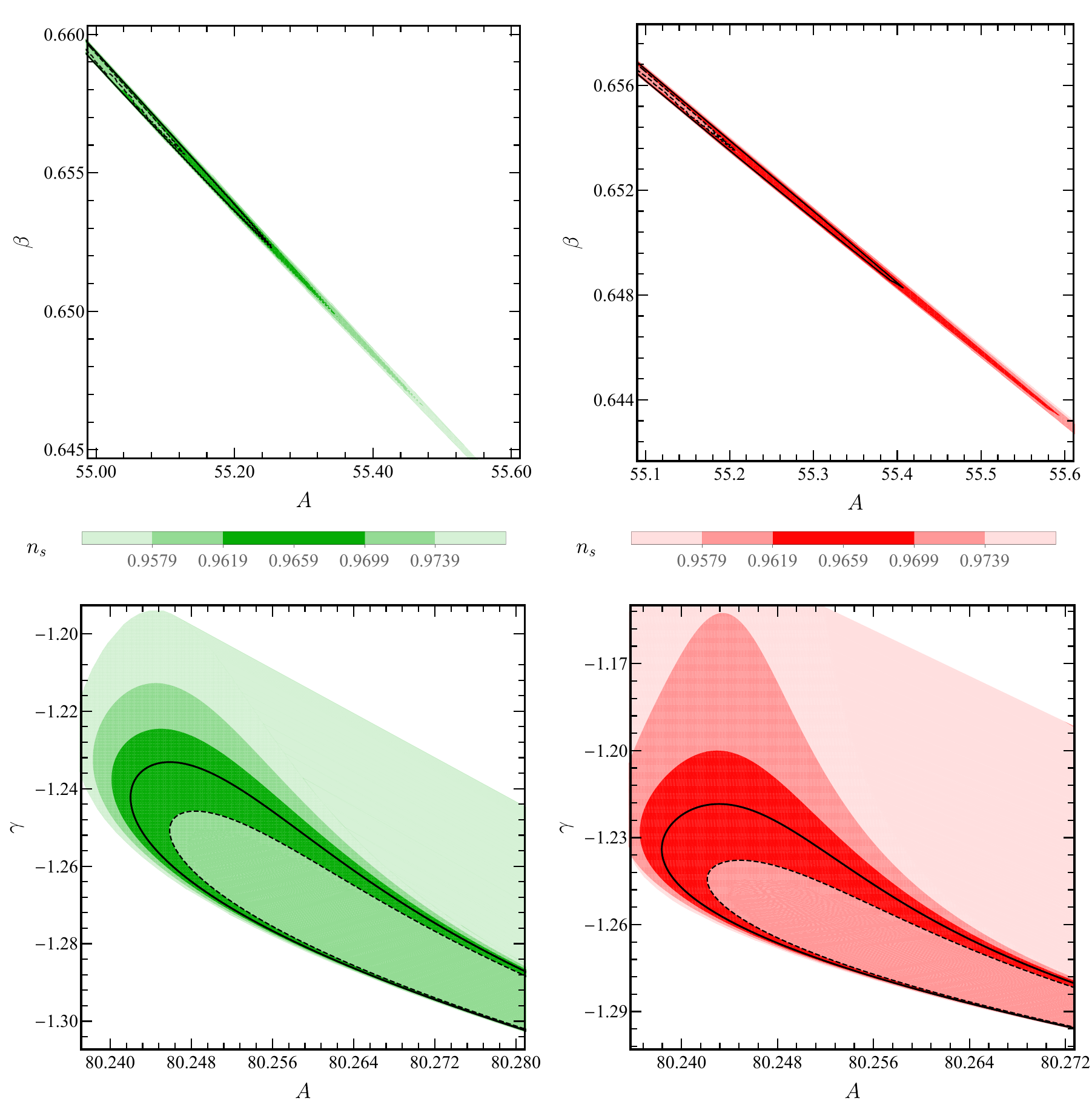}
\caption{The contour plot of the spectral index $n_s$ in the parameter planes $A-\beta$ for $\gamma=0$ (upper panels) and $A-\gamma$ for $\beta=0$ (lower panels). The
e-folds are taken to be 50 and 60 in the left panels and right panels respectively. The dark green regions on the left and dark red regions on the right stand for the $68\%$ CL region of $n_s$ for different values of e-folding. The black solid and dashed lines correspond to the central values of $n_s$ from Planck~\cite{Planck:2018jri} and ACT DR$6$~\cite{ACT:2025fju} respectively.}
\label{fig:paraPanel}
\end{figure}

In order to see the dependence of the inflation observables on the model parameters, we display the contour plot of $n_s$ in the $A-\gamma$ plane for $\beta=0$ and $A-\beta$ plane for $\gamma=0$ in figure~\ref{fig:paraPanel}, two values of $e$-folds $N_e= 50$ and $N_e= 60$ are adopted for illustration. The central values of $n_s$ from Planck~\cite{Planck:2018jri} and ACT DR$6$~\cite{ACT:2025fju} are shown as black solid and dashed lines respectively. One sees that there is enough parameter space to be compatible with the experimental data. We don't plot the results of the tensor-to-scalar ratio $r$, since it is too small to be testable.

%%%%%%%%%%%%%%%%%%%%%%%%%%%%%%%%%%%%%%%%%%%%%%%%%%%%%%%%%%%%

Before closing this section, we note that the total energy scale of the inflaton potential depends on the overall pre-factor  $\Lambda$ in eq.~\eqref{eq:simplifiedscalarpotential}, which is a function the value of $A$. Larger $A$ corresponds to smaller $\Lambda$, implying a smaller inflaton mass parameter. For example, for BP1 and BP2, we find $m_\phi = 4.5\times10^{7}\,\textrm{GeV}$ and $m_\phi = 3.9\times10^{6}\,\textrm{GeV}$, respectively. This suggests that the inflaton mass changes with the value of $A$. If we insist on the monotonic inflation potential between $\tau=i$ and $\tau=\tau_0$, $A$ should be larger than $43$, which gives us an upper bound on inflaton mass $m_\phi <1.5 \times 10^8 \,\textrm{GeV}$. This bound only applies to the special form of $H$ function in eq.~(\ref{eq:Hfun}). We note that the inflaton mass could be larger in other inflationary scenarios based on modular symmetry~\cite{Casas:2024jbw,Kallosh:2024ymt,Kallosh:2024pat}.

%%%%%%%%%%%%%%%%%%%%%%%%%%%%%%%%%%%%%%%%%%%%%%%%%%%
\subsection{Reheating from modulus decay}
After inflation, the inflaton field generally oscillates around its minimum of the potential and eventually decays into other particles in the Standard Model, which then thermalize, leading to a thermal bath. This process is called reheating \cite{Kofman:1994rk, Allahverdi:2010xz, Amin:2014eta, Lozanov:2019jxc}.
The temperature at the end of reheating is referred to as the reheating temperature, which is the highest temperature\footnote{We note that the maximum temperature can exceed $T_{\textrm{rh}}$ in non-instantaneous reheating scenarios \cite{Giudice:2000ex}. It has also been shown that if reheating occurs via inflaton decays to heavy neutrinos, the temperature approaches a constant, causing the maximum temperature to be close to the reheating temperature \cite{Cosme:2024ndc}. }  of the
radiation dominated era, can be defined via
$H(\Trh) = \frac23 \Gamma_\phi$, where $\Gamma_\phi$ denotes the inflaton decay rate and $H(\Trh)$ denotes the Hubble
parameter at $T = \Trh$. This gives  \cite{Drees:2024hok} 
\begin{equation}\label{trh}
\Trh = \sqrt { \frac {2} {\pi} } \left( \frac {10} {g_\star} \right)^{1/4}
\sqrt {M_{\textrm{pl}} \, \Gamma_\phi}\,,
\end{equation}
where $g_\star$ is the number of relativistic degrees of freedom contribution to the total radiation energy density. It reads $g_\star=106.75$ in the Standard model and roughly doubles in the MSSM, $g_\star=228.75$. 

After inflation, the Hubble scale of the universe decreases significantly during the oscillation of the inflaton field. Specifically, the relevant energy scale becomes much smaller than the Planck scale. Therefore, in the next section, it is sufficient to work in the global SUSY limit and neglect SUSY breaking effects. On one hand, SUSY breaking is highly model-dependent; on the other hand, for the canonical choice of a SUSY breaking scale around $\mathcal{O}(1)~\textrm{TeV}$, the mass splitting between particles and sparticles is not expected to significantly affect our results. Additionally, the expansion of the universe can also break SUSY, at the scale of the Hubble parameter~\cite{Dine:1983ys,Bertolami:1987xb,Copeland:1994vg,Dvali:1995mj,Dine:1995uk}. As mentioned above, the Hubble scale during reheating is approximately equal to the decay width of the inflaton and does not exceed $\mathcal{O}(1)~\textrm{GeV}$, provided the reheating temperature remains below $10^9~\textrm{GeV}$. SUSY can also be broken by thermal effects, see~\cite{Allahverdi:2004ix} for a possible application in cosmology. These effects will also be small due to small Yukawa coupling in the neutrino sector. Hence, we will perform the calculation using the same mass for particles and the corresponding sparticles. 

\subsubsection{Decay channels and reheating temperature}

In this section we analyse the possible inflaton decay channels within the current setup, with which we aim to compute the reheating temperature. 

We first evaluate the couplings in the SM sectors, where one can obtain the three point and four point vertices by expanding the mass matrices and Yukawa terms around the minimum of the canonically normalized field $\phi$. Three point vertices arise from inflaton-(s)neutrino-(s)neutrino interaction, and their Lagrangian reads\footnote{We refer to Ref.~\cite{Dedes:2007ef} for a detailed discussion on sneutrino mass matrix and Ref.~\cite{Dreiner:2008tw} for Feynman rules in 2 component notation.}:

\begin{equation}\label{eq:twobodydecay}
    \mathcal{L} =  \frac{1}{2}\frac{\Lambda_N}{M_{\text{pl}}} \lambda_1^{ij} \phi N^c_i N^c_j + \frac{1}{2}\frac{\Lambda_N^2}{M_{\text{pl}}}  \lambda_2^{ij}  \phi \widetilde{N}^{c*}_i \widetilde{N}^{c}_j+\text{h.c.}\,,
\end{equation}
where $N^c_i$ is the right handed neutrino and $\tilde{N}^c_i$ is the right handed sneutrino. We work in the basis where right handed neutrinos mass matrix is diagonal, and the coefficient matrices $\lambda_{1,2}^{ij}$ for benchmark values of the free parameters in eq.~\eqref{eq:parameters} read:

\begin{equation}
\lambda_1^{ij} =  \left[U^{T}_N \frac{d {\cal Y}_{N}}{d\phi} U_N\right]^{ij}\Bigg|_{\phi=\phi_0}=\left(
\begin{array}{ccc}
1.248+0.490i ~&~ 1.420i ~&~ -1.017 \\
 1.420i ~&~ -1.863+0.517i ~&~ 1.080i \\
- 1.017 ~&~ 1.080i ~&~ -0.616+1.006i \\
\end{array}
\right)\,,
\end{equation}
and 
\begin{equation}
\lambda_2^{ij}= \left[U^{\dagger}_N \frac{d ({\cal Y}_{N}^{\dag}{\cal Y}_{N})}{d \phi} U_N\right]^{ij}\Bigg|_{\phi=\phi_0}= \left(
\begin{array}{ccc}
3.423 ~&~ -0.106i ~&~ -4.262 \\
0.106i ~&~ -5.390 ~&~ -1.481i \\
-4.262 ~&~ 1.481i ~&~-3.470\\
\end{array}
\right)\,.
\end{equation}

The relevant two-body decay widths are given by
\begin{align}\label{eq:phiNN}
\Gamma({\phi\rightarrow N^c_i N^c_j})
= \frac{m_\phi}{8(1+\delta_{ij})\pi}\left(\frac{ \Lambda_N}{M_{\text{pl}}}\right)^2&\left(\left|\lambda_1^{ij}\right|^2 \left(1-\frac{M_i^2+M_j^2}{m_\phi^2}\right)-2\textrm{Re}[(\lambda_1^{ij})^2] \frac{M_i M_j}{m_\phi^2}\right)\nonumber \\&\times \sqrt{\left(1 - \frac{(M_j-M_i)^2}{m_\phi^2}\right)\left(1 - \frac{(M_j+M_i)^2}{m_\phi^2}\right)}\,,
\end{align}
\begin{align}\label{eq:phiNtildeNtilde}
\Gamma(\phi\to \widetilde{N}^c_i \widetilde{N}^{c*}_j) =\left|\frac{\Lambda_N^2\lambda_2^{ij}}{M_{\text{pl}}}\right|^2\frac{1}{16\pi m_\phi}\sqrt{ \left(1 - \frac{(M_i -M_j)^2}{m_\phi^2}\right) \left(1 - \frac{(M_i +M_j)^2}{m_\phi^2}\right) }\,,
\end{align}
where $M_i$ and $M_j$ denote the right handed neutrino masses. We note when $M_i +M_j  > m_\phi$ the two-body rates vanish, as expected due to the kinematic threshold. % \to

%%%%%%%%%%%%%%%%%%%%%%%%%%%%%%%%%%%%%%%%%%%%%%%%%%%%%%%%%%%%%

Analogously the Lagrangian relevant to the three body decay of inflaton is given by
\begin{align}\label{eq:3bodyL}
\nonumber \mathcal{L} = &\frac{\lambda_{3}^{ij}}{M_{\text{pl}}}  \phi N^c_i (L_j \cdot H_u) + \frac{\lambda_{3}^{ij}}{M_{\text{pl}}}  \phi \widetilde{N}^c_i (L_j \cdot \widetilde{H}_u)+ \frac{\lambda_{3}^{ij}}{M_{\text{pl}}}  \phi N^c_i (\widetilde{L}_j \cdot \tilde{H}_u)\\&+\frac{\lambda_4^{ij}   \Lambda_N}{M_{\text{pl}}} \phi \widetilde{N}_i^{c*}(\widetilde{L}_j \cdot H_u)+\text{h.c.}  \,,
\end{align}
where $L_j=\left(\nu_j\,,l_j\right)^T$, $H_u=\left(H^{+}_u\,,H^{0}_u\right)^T$ are $SU(2)$ doublet and $(L_j\cdot H_u) = \nu_j H^0_u - l_j H_u^{+}$. The contraction is the same for slepton $\tilde{L}$ and higgsino $\tilde{H}$. The coefficients matrices in eq.~(\ref{eq:3bodyL}) read:
\begin{equation}
\lambda_{3}^{ij} = \left[U^{T}_N \frac{d {\cal Y}_{D} }{d \phi} U_L^\nu\right]^{ij}\Bigg|_{\phi=\phi_0}= 
\left(
\begin{array}{ccc}
-0.490+1.248i ~&~ -1.347-0.124i ~&~- 0.003+0.748i \\
-1.492+0.124i ~&~ -0.517-1.864i ~&~ 0.830  \\
-0.003- 1.288i ~&~ -1.328-0.001i ~&~ 1.006+0.616i \\
\end{array}
\right)g_1\,,
\end{equation}
\begin{equation}
\lambda^{ij}_4 =  \left[U^{\dag}_N \frac{d ({\cal Y}_{N}^{\dag} {\cal Y}_{D}) }{d \phi} U_L^\nu\right]^{ij}\Bigg|_{\phi=\phi_0}=
\left(
\begin{array}{ccc}
2.921i ~&~ -0.225   ~&~ 3.902i  \\
-0.849 ~&~ -4.896i ~&~ -1.831 \\
-5.404i ~&~ -1.827 ~&~ 3.462i \\
\end{array}
\right)g_1\,.
\end{equation}
The decay width reads (see Appendix~\ref{sec:3-body} for details):
\begin{align}\label{eq:phiNHL}
    \Gamma(\phi \to  N^c_i (L_j \cdot H)  ) &= \Gamma(\phi \to  N^c_i(\widetilde{L}_j \cdot \widetilde{H}))\nonumber \\
    &= 2\times \left|\frac{\lambda_{3}^{ij}}{M_{\text{pl}}} \right|^2\frac{m_\phi^3}{768 \pi ^3} \left[1-6 \mu_N+3 \mu_N^2+2 \mu_N^3-6 \mu_N^2 \log(\mu_N)\right]\,,
\end{align}
where we use $(L_j \cdot H)$ to indicate there are two possible final states $\nu_j H^0_u$ or $l_j H_u^{+}$. The factor 2 accounts for two terms in $SU(2)$ contraction. It is also possible that inflaton decays into sneutrino, and the rates are
\begin{align}\label{eq:phiNHLtilde}
    \Gamma(\phi \to \widetilde{N}^c_i  (L_j \cdot \widetilde{H}) )  &= 2\times \left|\frac{\lambda_{3}^{ij}}{M_{\text{pl}}} \right|^2 \nonumber \\
    & \times \frac{m_\phi^3}{768 \pi ^3} \left[ 1+9 \mu_N-9 \mu_N^2-\mu_N^3+6 \mu_N \log(\mu_N)+6 \mu_N^2 \log(\mu_N)
    \right]\,;
\end{align}
\begin{align}\label{eq:phiNtildeHLtilde}
    \Gamma(\phi \to \widetilde{N}^c_i (\widetilde{L}_j \cdot H) ) & = 2\times \left|\frac{\lambda_4^{ij} \Lambda_N}{M_{\text{pl}}} \right|^2 \times \frac{m_\phi}{512 \pi ^3} \left[ 1 -\mu_N^2+2 \mu_N \log(\mu_N)\right]\,,
\end{align}
where $\mu_N = M_i^2 /{m_\phi^2}$. The three-body decay rate approaches zero when $\mu_N \to 1$, which is expected, as the decay channel becomes kinematically blocked in this case. 

Furthermore, we would like to mention that there is no effective interactions among the modulus $\tau$ and pure Higgs fields in our model, since both Higgs fields $H_u$ and $H_d$ are invariant under modular symmetry with zero modular weight. As a result, decay mode of $\phi$ into two Higgs is forbidden. Analogously the vector superfields are also modular invariant and their modular weights are vanishing. Therefore the effective operator $\phi F_{\mu\nu}\tilde{F}^{\mu\nu}$ can not be generated if the modular symmetry anomaly is cancelled for proper modular weights of quark and lepton fields~\cite{Feruglio:2024ytl}. Thus the decay of $\phi$ into two gauge bosons would be also forbidden.

By comparing the decay rates, we find that the channel in which the inflaton decays into two right-handed neutrinos (i.e. eq.~\eqref{eq:phiNN}) dominates in the regime $\mathcal{O}(10^{2}) \left(\frac{m_\phi}{10^{10}\,\text{GeV}} \right)^2~\text{GeV} < M_1 < m_\phi/2$. For $M_1 < \mathcal{O}(10^{2}) \left(\frac{m_\phi}{10^{10}\,\text{GeV}} \right)^2~\text{GeV}$ and $m_\phi/2 < M_1 < m_\phi$, the three-body channels eq.~\eqref{eq:phiNHL} and   eq.~\eqref{eq:phiNHLtilde} dominate. The decay widths discussed above are suppressed by the ratio of the inflaton mass to the Planck mass. Consequently, if the inflaton decays only into Standard Model particles, the reheating temperature remains relatively low. 

\begin{figure}[t!]
\def\sepf{0.8}
\centering
\includegraphics[scale=\sepf]{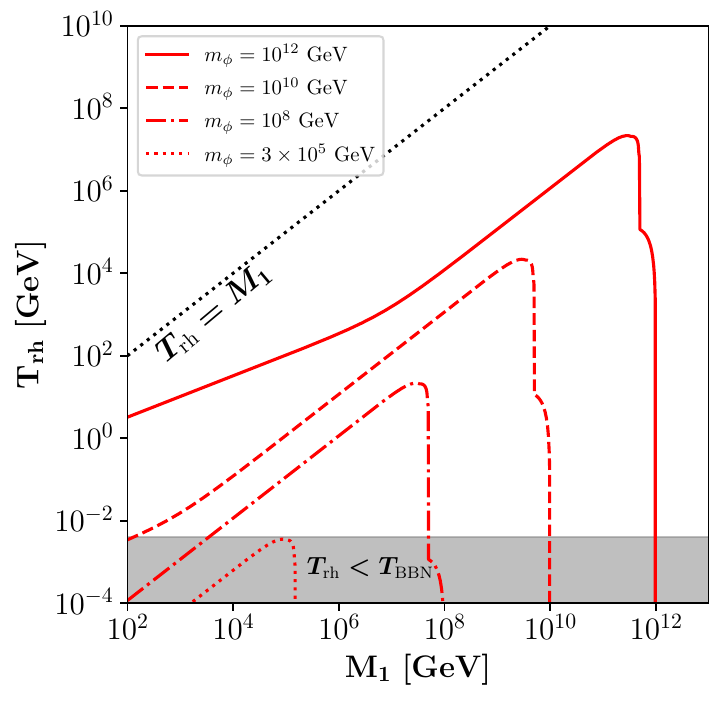}
\caption{Reheating temperature as function  of the lightest right handed neutrino mass $M_1$ and inflaton mass $m_\phi$ by considering inflaton two and three body decays.}
\label{fig:Trh}
\end{figure}
%%%%%%%%%%%%%%%%%%%%%%%%%%%%%%%%%%%%%%%%%%%%%%%

In Fig.~\ref{fig:Trh}, we show the reheating temperature as a function of the lightest right-handed neutrino mass, $M_1$, for various inflaton masses: $m_\phi = 10^{12}$ GeV (solid red), $m_\phi = 10^{10}$ GeV (dashed red), $m_\phi = 10^8$ GeV (dash-dotted red), and $m_\phi = 3 \times 10^5$ GeV (dotted red). These results are obtained by summing over 54 channels from the two- and three-body decays discussed earlier. Larger inflaton masses correspond to higher decay rates, resulting in larger reheating temperatures (cf. eq.~\eqref{trh}). This explains why the solid red line with $m_\phi = 10^{12}$ GeV lies above the lines for smaller inflaton masses. 

As mentioned above, in the regime $M_1 < \mathcal{O}(10^{2}) \left(\frac{m_\phi}{10^{10}\,\text{GeV}} \right)^2~\text{GeV}$, the three-body channels, eqs.~\eqref{eq:phiNHL} and \eqref{eq:phiNHLtilde}, dominate, resulting in $\Trh \propto \sqrt{M_1}$. Beyond this regime, $\Trh \propto M_1$ due to the dominance of the two-body rate, eq.~\eqref{eq:phiNN}. This explains the change in slope of the red lines when $M_1 \simeq  \mathcal{O}(10^{2}) \left(\frac{m_\phi}{10^{10}\,\text{GeV}} \right)^2~\text{GeV}$. The change in slopes is evident when compared to the reference black dotted line, where $\Trh =M_1$. It is also evident that, within the current setup, the reheating temperature remains below the mass of the lightest right-handed neutrino, implying that thermalization is strongly suppressed. We also note that the red lines feature a kink as $M_1 \to m_\phi/2$, where the three-body decay becomes dominant in this region.

To preserve the successful predictions of Big Bang Nucleosynthesis (BBN), it is required that $T_{\text{rh}} > 4\,\text{MeV}$~\cite{Kawasaki:2000en, Hannestad:2004px, deSalas:2015glj, Hasegawa:2019jsa}. Therefore, $T_{\text{rh}} < 4\,\text{MeV}$ is disallowed, as indicated by the gray region in Fig.~\ref{fig:Trh}. For the current model setup, we find that the inflaton mass must satisfy $m_\phi \gtrsim 3 \times 10^5~\text{GeV}$ to be consistent with BBN constraints. We note that our inflationary setup satisfies this condition as discussed at the Sec.~\ref{sec:modular_inflation}.

Given that the inflaton dominantly decays to right-handed neutrinos, we also investigate the possibility of realizing baryogenesis via non-thermal leptogenesis in Appendix~\ref{app:leptogenesis}. We find that achieving this within the current small-field hilltop inflationary framework is highly challenging due to the suppressed inflaton mass and low reheating temperature. A large-field inflationary scenario may provide a more viable solution. We leave this for a future work.

%%%%%%%%%%%%%%%%%%%%%%%%%%%%%%%%%%%%%%%%%%%%%%%%%%%%%%%

\section{Conclusion}\label{sec:conclusion}

In this work, we present a minimal model that attempts to simultaneously address the lepton flavor puzzle, inflation and post-inflationary reheating based on modular symmetry. We show that all the three aspects can be achieved collectively through the modulus field, without the need to introduce any additional new physics.

In the lepton sector, we employ a Type-I seesaw modular $A_4$ model to explain the smallness of neutrino masses. By assigning the standard mode (SM) fields and right handed neutrinos (RHNs) different modular weights and irreducible representations, modular symmetry determines the possible forms of the Yukawa interactions. After the modulus field acquires a VEV, modular symmetry is broken, and the Yukawa coefficients become fixed. We find that the VEV $\tau_0 = -0.484747 + 0.874655 i$, located around the fixed point $\omega=e^{i2\pi/3}$, can successfully reproduce SM observations in the lepton sectors as demonstrated in Sec.~\ref{sec:A4}.

We show that the same scalar potential that fixes the VEV of the modulus field can also account for inflation. To this end, this scalar potential must be sufficiently flat in a certain region. We consider inflation occurring around the fixed point $\tau = i$ and inflaton  oscillating at $\tau = \tau_0$, which can be realized with an appropriate superpotential, as demonstrated in Sec.~\ref{sec:inflation}. In this setup, the inflationary trajectory follows the arc of the fundamental domain, as shown in Fig.~\ref{fig:crosssecv2}, where the special properties of modular symmetry are maximally pronounced. Consequently, the inflationary scenario is similar to the hilltop model. We find that the model can perfectly fit the CMB observations, featuring a very small tensor-to-scalar ratio $r \sim \mathcal{O}(10^{-7})$. 

Although $r$ is too small to be testable, our model could be falsified through the measurement of the running of the spectral index $\alpha \equiv \mathrm{d}n_s / \mathrm{d}\ln k$, which falls in the range of $-\mathcal{O}{(10^{-4})}$ to $-\mathcal{O}{(10^{-3})}$ in our current setup.  The next generation of CMB experiments such as CMB-S4~\cite{Abazajian:2019eic}, and LiteBIRD~\cite{Matsumura:2013aja,LiteBIRD:2025trg}, combined with small-scale structure information (e.g., the Lyman-$\alpha$ forest), could probe $\alpha$ at the level of $10^{-4}$, making our inflation potential testable in future. A measurement consistent with our predictions would provide strong support for the modular-invariant inflation, underscoring its predictive power and constraining the viable parameter space to an even narrower region.

Any viable inflationary scenario must also explain how the Universe reheats. A novel feature of our setup is that the channels for post-inflationary reheating are automatically generated to explain the observations in the lepton sector. In particular, the expansion of the modular forms around the minimum gives rise to interactions between the inflaton and other particles, including the SM Higgs, leptons, and RHNs. After inflation, the inflaton decays through these channels, which can reheat the universe. We compute all relevant channels, including inflaton two- and three-body decays. We find that, due to the Planck-suppressed interactions, the reheating temperature tends to be low unless the inflaton mass is larger, as depicted in Fig.~\ref{fig:Trh}. The highest reheating temperature occurs when the RHN masses approach their kinematical threshold. Interestingly, we find a parameter space that yields a sufficiently high reheating temperature to preserve the successful predictions of Big Bang Nucleosynthesis (BBN). This requires the inflaton satisfy $m_\phi \gtrsim \mathcal{O}(10^5)~\text{GeV}$.

We further explore the possibility of explaining the baryon asymmetry of the Universe (BAU) via leptogenesis in the Appendix~\ref{app:leptogenesis}. We apply the non-thermal leptogenesis mechanism, as the temperature in our framework is lower than the RHN mass, implying that the thermal production of RHNs is Boltzmann-suppressed. We find that, in order to account for the observed BAU, the inflaton and the lightest RHN masses must satisfy $m_\phi \gtrsim \mathcal{O}(10^{11})~\text{GeV}$ and $M_1 \gtrsim \mathcal{O}(10^{11})~\text{GeV}$, as shown in Fig.~\ref{fig:ps_BAU}. 
We note that the small-field hilltop inflationary model considered in the current setup cannot satisfy this condition. An interesting direction is to explore other inflationary setups, such as those presented very recently in Refs.~\cite{Casas:2024jbw, Kallosh:2024ymt, Kallosh:2024pat}. Although our current setup does not fully account for the BAU, we believe that our approach provides a valuable basis for further exploration of post-inflationary cosmology within the framework of modular invariance.

\section*{Acknowledgments}
%%%%%%%%%%%%%%%%%%%%%%%%%%%%%%%%%%%%%%%%% 
WBZ would like to thank Professor Manuel Drees for fruitful discussions and carefully reading and commenting on the draft. GJD is grateful to Professors Serguey Petcov and Ye-Ling Zhou for helpful discussions. 
The authors thank the Mainz Institute for Theoretical Physics (MITP) for its hospitality during the workshop ``Modular Invariance Approach to the Lepton and Quark Flavor Problems: from Bottom-up to Top-down'', where this collaboration was initiated. GJD and SYJ are supported by the National Natural Science Foundation of China under Grant No. 12375104. YX has received support from the Cluster of Excellence ``Precision Physics, Fundamental Interactions, and Structure of Matter'' (PRISMA$^+$ EXC 2118/1) funded by the Deutsche Forschungsgemeinschaft (DFG, German Research Foundation) within the German Excellence Strategy (Project No. 390831469).

\begin{appendix}

\section{\label{app:A4-MF} Finite modular group $\Gamma_3\cong A_4$ and modular forms of level 3 }

The level 3 finite modular group $\Gamma_3$ is isomorphic to the  $A_4$ group which is the even permutation group of four objects, and it can be generated by the modular generators $S$ and $T$ satisfying the following relations
\begin{equation}
\label{eq:A4-rules}S^2=(ST)^3=T^3=1\,.
\end{equation}
The $\Gamma_3\cong A_4$ group has three singlet representations $\bm{1}$, $\bm{1}'$ and $\bm{1}''$, and one triplet representation $\bm{3}$. In the singlet representations, the generators $S$ and $T$ are represented by ordinary numbers. From the multiplication rules in eq.~\eqref{eq:A4-rules}, it is straightforward to obtain the singlet representations as follows,
\begin{eqnarray}
\nonumber && \bm{1}~:~ S=1\,,\qquad T=1 \,,  \\
\nonumber &&\bm{1}'~:~ S=1\,,\qquad T=e^{2\pi i/3}\equiv\omega \,,  \\
 &&\bm{1}''~:~S=1\,,\qquad T=e^{4\pi i/3}=\omega^{2} \,.
\end{eqnarray}
For the triplet representation $\bm{3}$, the generators $S$ and $T$ are represented by
\begin{equation}
\bm{3}~:~ S=\frac{1}{3}\begin{pmatrix}
-1~& 2  ~& 2  \\
 2  ~& -1  ~& 2 \\
2 ~& 2 ~& -1
\end{pmatrix}, \qquad
T=\begin{pmatrix}
1 ~&~ 0 ~&~ 0 \\
0 ~&~ \omega ~&~ 0 \\
0 ~&~ 0 ~&~ \omega^{2}
\end{pmatrix} \,.
\end{equation}
The tensor products of singlet representations are
\begin{equation}
\bm{1}'\otimes\bm{1}'=\bm{1}''\,,\quad \bm{1}''\otimes\bm{1}''=\bm{1}'\,,\quad 
\bm{1}'\otimes\bm{1}''=\bm{1}\,.
\end{equation}
The tensor product of two $A_4$ triplets is
\begin{equation}
\bm{3}\otimes\bm{3}=\bm{1}\oplus \bm{1}'\oplus\bm{1}''\oplus\bm{3}_S\oplus\bm{3}_A\,,
\end{equation}
where $\bm{3}_S$ and $\bm{3}_A$ denote the symmetric and antisymmetric triplet contractions respectively. In terms of the components of the two triplets $\bm{a}=\left(a_1, a_2, a_3\right)^{T}$ and $\bm{b}=\left(b_1, b_2, b_3\right)^{T}$, we have
\begin{eqnarray}
\nonumber&\left(\bm{a}\otimes\bm{b}\right)_{\bm{1}}=a_1b_1+a_2b_3+a_3b_2\,,\\
\nonumber&\left(\bm{a}\otimes\bm{b}\right)_{\bm{1}'}=a_1b_2+a_2b_1+a_3b_3\,,\\
\nonumber&\left(\bm{a}\otimes\bm{b}\right)_{\bm{1}''}=a_1b_3+a_2b_2+a_3b_1\,, \\
\label{eq:A4-CG-coef}&\left(\bm{a}\otimes\bm{b}\right)_{\bm{3}_S}=\begin{pmatrix}
2a_1b_1-a_2b_3-a_3b_2 \\
2a_3b_3-a_1b_2-a_2b_1\\
2a_2b_2-a_1b_3-a_3b_1
\end{pmatrix}\,,\quad \left(\bm{a}\otimes\bm{b}\right)_{\bm{3}_A}=\begin{pmatrix}
a_2b_3-a_3b_2\\
a_1b_2-a_2b_1\\
a_3b_1-a_1b_3
\end{pmatrix}\,.
\end{eqnarray}

\subsection{Modular forms of level 3}\label{app:modulargrpoup}

The even weight modular forms of level 3 can be arranged into multiplets of $A_4$ up to the automorphy factor. At weight $k=2$, there are only three linearly independent modular forms $Y_1(\tau)$, $Y_2(\tau)$ and $Y_3(\tau)$ which form a $A_4$ triplet $Y^{(2)}_{\bm{3}}(\tau)\equiv\left(Y_1(\tau)\,, Y_2(\tau) \,, Y_3(\tau)\right)^T$~\cite{Feruglio:2017spp}. One can express $Y_{1,2,3}(\tau)$ in terms of the product of Dedekind eta-function~\cite{Liu:2019khw} or its derivative~\cite{Feruglio:2017spp}. In practice, the first few terms of the $q$-expansion of $Y_{1,2,3}(\tau)$ provide sufficiently accurate approximation~\cite{Feruglio:2017spp},
\begin{eqnarray}
\nonumber&&Y_1(\tau)=1+12q+36q^2+12q^3+84q^4+72q^5+\ldots \,, \\
\nonumber&&Y_2(\tau)=-6q^{1/3}\left(1+7q+8q^2+18q^3+14q^4+31q^5+\ldots\right)\,, \\
&&Y_3(\tau)=-18q^{2/3}\left(1+2q+5q^2+4q^3+8q^4+6q^5+\ldots\right)\,.
\end{eqnarray}
Using the tensor product decomposition in eq.~\eqref{eq:A4-CG-coef},  the higher weight modular forms of level 3 can be written as polynomials of $Y_1(\tau)$, $Y_2(\tau)$ and $Y_3(\tau)$. At weight 4, the tensor product of $Y^{(2)}_{\bm{3}}\otimes Y^{(2)}_{\bm{3}}$ gives rise to three linearly independent modular multiplets,
\begin{eqnarray}
\nonumber Y^{(4)}_{\bm{1}}&=&(Y^{(2)}_{\bm{3}}\otimes Y^{(2)}_{\bm{3}})_{\bm{1}}=Y_1^2+2 Y_2 Y_3\,, \\
\nonumber Y^{(4)}_{\bm{1}'}&=&(Y^{(2)}_{\bm{3}}\otimes Y^{(2)}_{\bm{3}})_{\bm{1}'}=Y_3^2+2 Y_1 Y_2\,,\\
Y^{(4)}_{\bm{3}}&=&\frac{1}{2}(Y^{(2)}_{\bm{3}}\otimes Y^{(2)}_{\bm{3}})_{\bm{3}_S}=\left(\begin{array}{c}
Y_1^2-Y_2 Y_3\\
Y_3^2-Y_1 Y_2\\
Y_2^2-Y_1 Y_3
\end{array}\right)\,.
\end{eqnarray}
The weight 6 modular forms of level 3 decompose as $\mathbf{1}\oplus\mathbf{3}\oplus\mathbf{3}$ under $A_4$, there are two independent triplet modular forms and they can be chosen as
\begin{eqnarray}
Y^{(6)}_{\bm{1}}&=&\left(Y^{(2)}_{\bm{3}}\otimes Y^{(4)}_{\bm{3}}\right)_{\bm{1}}=Y_1^3+Y_2^3+Y_3^3-3 Y_1 Y_2 Y_3\,,\nonumber\\
Y^{(6)}_{\bm{3}I}&=&\left(Y^{(2)}_{\bm{3}}\otimes Y^{(4)}_{\bm{1}}\right)_{\bm{3}}=(Y_1^2+2Y_2Y_3)\left(\begin{array}{c}
Y_1\\
Y_2\\
Y_3
\end{array}\right)\,,\nonumber\\
\label{eq:weight6-MF}Y^{(6)}_{\bm{3}II}&=&\left(Y^{(2)}_{\bm{3}}\otimes Y^{(4)}_{\bm{1}'}\right)_{\bm{3}}=
(Y_3^2+2 Y_1Y_2)\left(\begin{array}{c}
Y_3\\
Y_1\\
Y_2
\end{array}\right)\,.
\end{eqnarray}
The Dedekind eta function $\eta(\tau)$, is a modular function of ``weight $1/2$'' defined as 
\begin{equation}
\label{eq:eta-func}\eta(\tau)=q^{1/24}\prod^{\infty}_{n=1}(1-q^n),\quad~~
q\equiv e^{2\pi i\tau}\,,
\end{equation}
which satisfies the following modular transformation identities: $\eta(\tau+1)=e^{i\pi/12}\eta(\tau)$ and $\eta(-1/\tau)=\sqrt{-i\tau}\eta(\tau)$.
The $q$-expansion of eta function is given by
\begin{equation}
\eta=q^{1/24}[1-q-q^2+q^5+q^7-q^{12}-q^{15}+{\cal O}(q^{22})]\,.
\end{equation}
The $j$ function is related to the Dedekind eta and its derivatives as follow,
\begin{equation}
j=\left(\frac{72}{\pi^2}\frac{\eta\eta''-\eta'^2}{\eta^{10}}\right)^3 = \left[\frac{72}{\pi^2 \eta^6}\left(\frac{\eta'}{\eta^3}\right)'\right]^3\,,
\end{equation}
where prime denotes derivative with respect to $\tau$.
\section{Vacuum structure of modulus}
\label{app:modular_derivative}
In this section, we exam the properties of $\tau=\omega$ and $\tau=\tau_0$ in details. Both are minima, but they have different potential values. For convenience, we denote
\begin{equation}
    \mathcal{P}(j(\tau)) = 1 +\beta \left(1-\frac{j(\tau)}{1728}\right)+\gamma \left(1-\frac{j(\tau)}{1728}\right)^2\,
\end{equation}
in eq.~\eqref{eq:Hfun}. For $\tau=\omega$, the potential and its first-order derivatives read
\begin{equation}
V(\omega)=\Lambda^4 \frac{(2\pi)^{12}}{3^3 \Gamma^{18}(1/3)}(A-3)|j^2(\tau_0){\cal P}(0)|^2\,,\quad \partial_\tau V(\omega) = \partial_{\bar{\tau}}V(\omega)=0\,,
\end{equation}
implying $V(\omega)$ will be positive-definite as long as $A > 3$.  As $\omega$ is a fix point under modular transformation, Modular symmetry ensures the first-order derivatives to vanish. The second-order derivatives of the potential, forming the Hessian matrix, are
\begin{equation}
\frac{\partial^2 V}{\partial \rho^2}\Bigg|_{\tau=\omega}=\frac{\partial^2 V}{\partial \theta^2}\Bigg|_{\tau=\omega}=2\Lambda^4\frac{(2\pi)^{12}}{3^3\Gamma^{18}(1/3)}\left(A-2\right)|j^2(\tau_0){\cal P}(0)|^2\,,\quad \frac{\partial^2 V}{\partial \theta \partial \rho}\Bigg|_{\tau=\omega} = 0\,.
\end{equation}
Because the Hessian matrix is positive-definite, $\tau=\omega$ is a local minimum.

Unlike $\omega$, the property at $\tau_0$
heavily rely on the special form of $H(\tau)$. Using that $H(\tau_0)=\partial_\tau H(\tau_0)=0$, the potential and its first-order derivatives are
\begin{equation}
V(\tau_0) = 0 \,,\quad \partial_\tau V(\tau_0) = \partial_{\bar{\tau}}V(\omega)= 0\,.
\end{equation}
and the second-order derivatives read
\begin{equation}
\frac{\partial^2 V}{\partial \rho^2}\Bigg|_{\tau=\tau_0}=\frac{\partial^2 V}{\partial \theta^2}\Bigg|_{\tau=\tau_0}=\frac{4\left|(\partial_\tau j(\tau_0))^2{\cal P}(j(\tau_0))\right|^2}{3\sin(\text{arg}(\tau_0))\abs{\eta(\tau_0)}^{12}}>0\,,\quad \frac{\partial^2 V}{\partial \theta \partial \rho}\Bigg|_{\tau=\tau_0}= 0\,.
\end{equation}
In this case, the Hessian matrix is positive-definite when $\tau_0$ stays in the fundamental domain. Since our potential is semi-positive, the vanishing potential at $\tau_0$ ensures it is a global minimum. 

The property at $\tau=i$ is rather non-trivial. In this paper we need $i$ to be a saddle point, which is not a general conclusion. The Hessian matrix now depends on the parameters in $\mathcal{P}(j(\tau))$. Thus, we only show the numerical result in Fig.~\ref{fig:crosssecv2}.

%%%%%%%%%%%%%%%%%%%%%%%%%%%%%%%%%%%%%%%%%%%%
\section{Inflaton decay rates}\label{sec:3-body}
In this section we will present calculations about inflaton decays. As we are dealing with Majorana fermions, this calculation will be carried out in the two component notations.
\subsection{Inflaton 2-body decay}
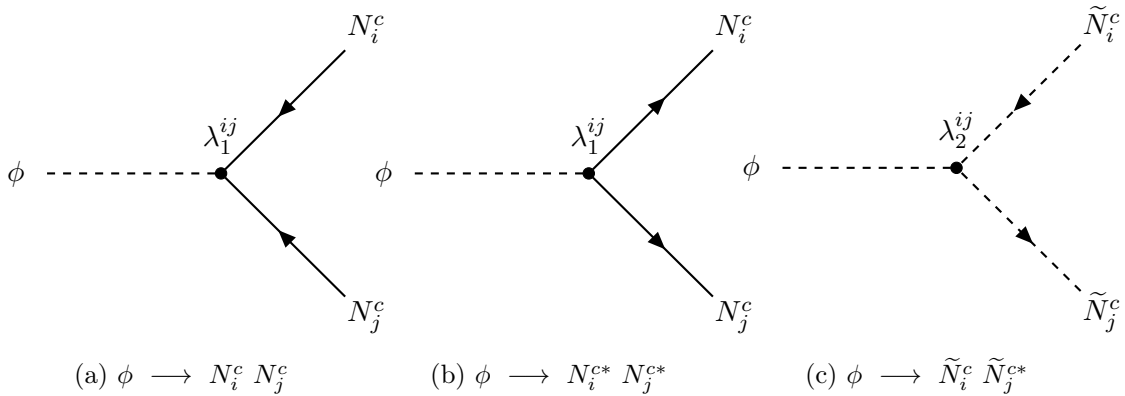
\begin{figure}
\begin{subfigure}{.32\textwidth}
    \centering
\begin{tikzpicture}
\begin{feynman}
			\vertex[dot] at (0,0)(o){};
			\vertex[] at (45:2.5)(e1){};
			\vertex[] at (315:2.5)(e2){};
			\vertex[] at (180:2.5)(e3){};
			\vertex[] at ($(e1)+(45:0.2)$)(t1){$N^c_i$};
			\vertex[] at ($(e2)+(315:0.2)$)(t2){$N^c_j$};
			\vertex[] at ($(e3)+(180:0.2)$)(t3){$\phi$};
			\vertex[] at (0,0.5)(t4){$\lambda^{ij}_1$};
			\diagram*{
				(e1)--[thick,fermion](o),
				(e2)--[thick,fermion](o)--[thick,scalar](e3)};
\end{feynman}
\end{tikzpicture}
    \caption{$\phi~\longrightarrow~N^c_i~N^c_j$}
\end{subfigure}%
\begin{subfigure}{.32\textwidth}
    \centering
\begin{tikzpicture}
\begin{feynman}
			\vertex[dot] at (0,0)(o){};
			\vertex[] at (45:2.5)(e1){};
			\vertex[] at (315:2.5)(e2){};
			\vertex[] at (180:2.5)(e3){};
			\vertex[] at ($(e1)+(45:0.2)$)(t1){$N^c_i$};
			\vertex[] at ($(e2)+(315:0.2)$)(t2){$N^c_j$};
			\vertex[] at ($(e3)+(180:0.2)$)(t3){$\phi$};
			\vertex[] at (0,0.5)(t4){$\lambda^{ij}_1$};
			\diagram*{
				(e1)--[thick,anti fermion](o),
				(e2)--[thick,anti fermion](o)--[thick,scalar](e3)};
\end{feynman}
\end{tikzpicture}
    \caption{$\phi~\longrightarrow~N^{c*}_i~N^{c*}_j$}
\end{subfigure}%
\begin{subfigure}{.32\textwidth}
    \centering
\begin{tikzpicture}
\begin{feynman}
			\vertex[dot] at (0,0)(o){};
			\vertex[] at (45:2.5)(e1){};
			\vertex[] at (315:2.5)(e2){};
			\vertex[] at (180:2.5)(e3){};
			\vertex[] at ($(e1)+(45:0.2)$)(t1){$\widetilde{N}^{c}_i$};
			\vertex[] at ($(e2)+(315:0.2)$)(t2){$\widetilde{N}^c_j$};
			\vertex[] at ($(e3)+(180:0.2)$)(t3){$\phi$};
			\vertex[] at (0,0.5)(t4){$\lambda^{ij}_2$};
			\diagram*{
				(e1)--[thick,charged scalar](o),
				(e2)--[thick,anti charged scalar](o)--[thick,scalar](e3)};
\end{feynman}
\end{tikzpicture}
\caption{$\phi~\longrightarrow~\widetilde{N}^c_i~\widetilde{N}^{c*}_j$}
\end{subfigure}
\caption{Feynman diagrams for inflaton two-body decay.}
\end{figure}

We consider the inflaton two body decay in the following Lagrangian:
\begin{equation}
    \mathcal{L} =  \frac{1}{2}\frac{\Lambda_N}{M_{\text{pl}}} \lambda_1^{ij} \phi N^c_i N^c_j + \frac{1}{2}\frac{\Lambda_N^2}{M_{\text{pl}}}  \lambda_2^{ij}  \phi \widetilde{N}^{c*}_i \widetilde{N}^{c}_j+\text{h.c.}\,,
\end{equation}
where $N^c_i$ is a Majorana particle with mass $M_i$ and $\widetilde{N}^{c}_i$ is corresponding sfermion. We first consider inflaton decay to two fermions, in the two component notation, matrix element reads:
\begin{equation}
i{\cal M} = y(\vec{p}_1,s_1)^\alpha (i\frac{ \Lambda_N}{M_{\text{pl}}}\lambda_1^{ij}\delta_\alpha{}^\beta)y(\vec{p}_2,s_2)_\beta + x^\dag(\vec{p}_1,s_1)_{\dot{\alpha}} (i\frac{ \Lambda_N}{M_{\text{pl}}}\lambda^{ij*}_{1}\delta^{\dot{\alpha}}{}_{\dot{\beta}})x^\dag(\vec{p}_2,s_2)^{\dot{\beta}}\,,
\end{equation}
where $x,y$ are the two component spinor wave functions, which play the same role as $u,v$ in four component notation. After taking hermitian conjugate and performing spin sum, we have:
\begin{eqnarray}\label{eq:Msqfermion}
\nonumber |{\cal M}|^2 &=& 4 \left(\frac{ \Lambda_N}{M_{\text{pl}}}\right)^2\left(| \lambda_1^{ij}|^2 p_i\cdot p_j - \text{Re}\left[(\lambda_1^{ij})^2\right] M_iM_j\right)\\
&=& 4\left(\frac{ \Lambda_N}{M_{\text{pl}}}\right)^2\left(|\lambda_1^{ij}|^2\frac{m^2_\phi - (M^2_j+M_i^2)}{2 }  - \text{Re}\left[(\lambda_1^{ij})^2\right] M_iM_j\right)\,,
\end{eqnarray}
where $M_i,M_j$ are mass of $N^c_i,N^c_j$, respectively.
For inflation decay to two scalars, the matrix element is much simpler:
\begin{equation}
|{\cal M}|^2 =  \left|\frac{\Lambda_N^2}{M_{\text{pl}}}  \lambda_2^{ij}\right|^2\,,
\end{equation}
The phase space integral can be performed as usual, the total decay rates are:

\begin{align}
\Gamma(\phi\to \widetilde{N}^c_i \widetilde{N}^{c*}_j) =\left|\frac{\Lambda_N^2\lambda_2^{ij}}{M_{\text{pl}}}\right|^2\frac{1}{16\pi m_\phi}\sqrt{ \left(1 - \frac{(M_i -M_j)^2}{m_\phi^2}\right) \left(1 - \frac{(M_i +M_j)^2}{m_\phi^2}\right) }\,,
\end{align}
for inflaton decay to two sfermions and:
\begin{align}
\Gamma({\phi\rightarrow N^c_i N^c_j})
= \frac{m_\phi}{8(1+\delta_{ij})\pi}\left(\frac{\Lambda_N}{M_{\text{pl}}}\right)^2&\left(\left|\lambda_1^{ij}\right|^2 \left(1-\frac{M_i^2+M_j^2}{m_\phi^2}\right)-2\textrm{Re}[(\lambda_1^{ij})^2] \frac{M_i M_j}{m_\phi^2}\right)\nonumber \\&\times \sqrt{\left(1 - \frac{(M_j-M_i)^2}{m_\phi^2}\right)\left(1 - \frac{(M_j+M_i)^2}{m_\phi^2}\right)}\,,
\end{align}
for inflaton decay to two fermions. Note $\delta_{ij}$ accounts for the effects of identical particles when $i=j$.

\subsection{Inflaton 3-body decay}
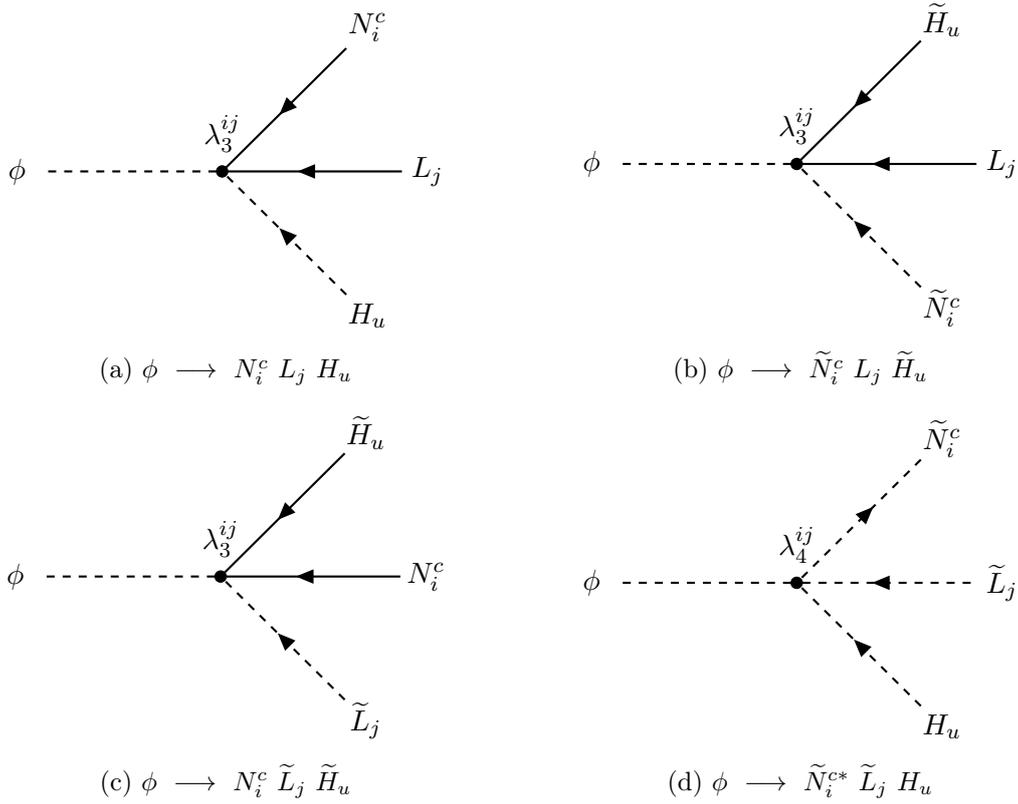
\begin{figure}
\begin{subfigure}{.5\textwidth}
    \centering
\begin{tikzpicture}
\begin{feynman}
			\vertex[dot] at (0,0)(o){};
			\vertex[] at (45:2.5)(e1){};
			\vertex[] at (0:2.5)(e2){};
			\vertex[] at (315:2.5)(e3){};
			\vertex[] at (180:2.5)(e4){};
			\vertex[] at ($(e1)+(45:0.2)$)(t1){$N^c_i$};
			\vertex[] at ($(e2)+(0:0.2)$)(t2){$L_j$};
			\vertex[] at ($(e3)+(315:0.2)$)(t3){$H_u$};
			\vertex[] at ($(e4)+(180:0.2)$)(t4){$\phi$};
			\vertex[] at (0,0.5)(t5){$\lambda^{ij}_3$};
			\diagram*{
				(e1)--[thick,fermion](o)--[thick,anti fermion](e2),
				(e3)--[thick,charged scalar](o)--[thick,scalar](e4)};
\end{feynman}
\end{tikzpicture}
    \caption{$\phi~\longrightarrow~N^c_i~L_j~H_u$}\label{fig:3-body1}
\end{subfigure}%
\hfill
\begin{subfigure}{.5\textwidth}
    \centering
\begin{tikzpicture}
\begin{feynman}
	\vertex[dot] at (0,0)(o){};
			\vertex[] at (45:2.5)(e1){};
			\vertex[] at (0:2.5)(e2){};
			\vertex[] at (315:2.5)(e3){};
			\vertex[] at (180:2.5)(e4){};
			\vertex[] at ($(e1)+(45:0.2)$)(t1){$\widetilde{H}_u$};
			\vertex[] at ($(e2)+(0:0.2)$)(t2){$L_j$};
			\vertex[] at ($(e3)+(315:0.2)$)(t3){$\widetilde{N}^c_i$};
			\vertex[] at ($(e4)+(180:0.2)$)(t4){$\phi$};
			\vertex[] at (0,0.5)(t5){$\lambda^{ij}_3$};
			\diagram*{
				(e1)--[thick,fermion](o)--[thick,anti fermion](e2),
				(e3)--[thick,charged scalar](o)--[thick,scalar](e4)};
\end{feynman}
\end{tikzpicture}
\caption{$\phi~\longrightarrow~\widetilde{N}^c_i~L_j~\widetilde{H}_u$}\label{fig:3-body2}
\end{subfigure}%

\medskip
\begin{subfigure}{.5\textwidth}
    \centering
\begin{tikzpicture}
\begin{feynman}
			\vertex[dot] at (0,0)(o){};
			\vertex[] at (45:2.5)(e1){};
			\vertex[] at (0:2.5)(e2){};
			\vertex[] at (315:2.5)(e3){};
			\vertex[] at (180:2.5)(e4){};
			\vertex[] at ($(e1)+(45:0.2)$)(t1){$\widetilde{H}_u$};
			\vertex[] at ($(e2)+(0:0.2)$)(t2){$N^c_i$};
			\vertex[] at ($(e3)+(315:0.2)$)(t3){$\widetilde{L}_j$};
			\vertex[] at ($(e4)+(180:0.2)$)(t4){$\phi$};
			\vertex[] at (0,0.5)(t5){$\lambda^{ij}_3$};
			\diagram*{
				(e1)--[thick,fermion](o)--[thick,anti fermion](e2),
				(e3)--[thick,charged scalar](o)--[thick,scalar](e4)};
\end{feynman}
\end{tikzpicture}
    \caption{$\phi~\longrightarrow~N^c_i~\widetilde{L}_j~\widetilde{H}_u$} \label{fig:3-body3}
\end{subfigure}%
\hfill
\begin{subfigure}{.5\textwidth}
    \centering
\begin{tikzpicture}
\begin{feynman}
			\vertex[dot] at (0,0)(o){};
			\vertex[] at (45:2.5)(e1){};
			\vertex[] at (0:2.5)(e2){};
			\vertex[] at (315:2.5)(e3){};
			\vertex[] at (180:2.5)(e4){};
			\vertex[] at ($(e1)+(45:0.2)$)(t1){$\widetilde{N}^{c}_i$};
			\vertex[] at ($(e2)+(0:0.2)$)(t2){$\widetilde{L}_j$};
			\vertex[] at ($(e3)+(315:0.2)$)(t3){$H_u$};
			\vertex[] at ($(e4)+(180:0.2)$)(t4){$\phi$};
			\vertex[] at (0,0.5)(t5){$\lambda^{ij}_4$};
			\diagram*{
				(e1)--[thick,anti charged scalar](o)--[thick,anti charged scalar](e2),
				(e3)--[thick,charged scalar](o)--[thick,scalar](e4)};
\end{feynman}
\end{tikzpicture}
\caption{$\phi~\longrightarrow~\widetilde{N}^{c*}_i~\widetilde{L}_j~H_u$} \label{fig:3-body4}
\end{subfigure}
\caption{Feynman diagrams for inflaton three-body decay.}
\end{figure}
For three-body decays, there are four channels as shown in Fig.~\ref{fig:3-body1}  -Fig.~\ref{fig:3-body4}. We first consider the inflaton three body decay $\phi(p) \to H(k_1) L(k_2) N^c(k_3)$, namely the process shown in Fig.~\ref{fig:3-body1}. The relevant Lagrangian is:
\begin{equation}
\mathcal{L} =   \frac{\lambda_{3}^{ij}}{M_{\text{pl}}}  \phi N^c_i (L_j \cdot H_u) +h.c.\,,
\end{equation}
where $L_j=\left(\nu_j\,,l_j\right)^T$, $H_u=\left(H^{+}_u\,,H^{0}_u\right)^T$ are SU(2) doublet and $(L_j\cdot H_u) = \nu_j H^0_u - l_j H^{+}$. 

In the following, we will neglect the Higgs and light neutrino masses, as they are much smaller compared to the inflaton mass and the right-handed neutrino mass.
The spin summed, squared matrix element for a single combination ($\nu_j H^0_u$ or $l_j H^{+}$) reads:
\begin{equation}
\abs{\mathcal{M}}^2=\abs{\frac{\lambda_{3}^{ij}}{M_{\text{pl}}} }^2 4 (k_2\cdot k_3)\,,
\end{equation}
With the squared matrix element, we can further compute the three-body decay rate, which is:
\begin{align}\label{eq:3-body}
\Gamma(\phi \to  N^c_i (L_j \cdot H)  ) \equiv \frac{1}{2\, m_\phi}\int d\Pi_3 \abs{\mathcal{M}}^2\,.
\end{align}
The three--body phase space integral can  be further  written as (see e.g. Sec. 20 of Ref.~\cite{Schwartz:2014sze} or step-by-step computation in Appendix C of Ref.~\cite{Xu:2022qpx})
\begin{equation}\label{eq:ps_integral}
\int d\Pi_3  =  \frac{m_\phi^2}{128\pi^3} \int^{1-\mu_N}_0 d x_1
\int^{1-\frac{\mu_N}{1-x_1}}_{1-x_1-\mu_N} d x_2\,,
\end{equation}
where $x_i = 2 E_i / m_\phi\,, i = 1,2$, where $E_1$ is the energy of the Higgs boson, and $E_2$ is the energy of the charged lepton in the inflaton rest frame, with $\mu_N \equiv M_N^2 / m_\phi^2$. We have neglected all final state masses except for that of the right-handed neutrino (RHN).
Using eq.~\eqref{eq:ps_integral}, we find that the three-body decay rate eq.~\eqref{eq:3-body} becomes 
\begin{align}
\Gamma(\phi \to  N^c_i (L_j \cdot H)  ) = 2\times \left|\frac{\lambda_{3}^{ij}}{M_{\text{pl}}} \right|^2\frac{m_\phi^3}{768 \pi ^3} \left[1-6 \mu_N+3 \mu_N^2+2 \mu_N^3-6 \mu_N^2 \log(\mu_N)\right]\,.
\end{align}
where  we use 2 to count 2 possible combinations in the $SU(2)$ contraction. We note that in the limit $\mu_N \to 1$, the rate $\Gamma_{\phi \to HLN} \to 0$, as expected, since the decay becomes kinematically blocked in this scenario. 

For other channels, the procedure is similar. In particular for  $\phi(p) \to \widetilde{H}(k_1) L(k_2) \widetilde{N}^c(k_3)$ (Fig.~\ref{fig:3-body2}), we find the squared matrix element is given by 
\begin{equation}
\abs{\mathcal{M}}^2=\abs{\frac{\lambda_{3}^{ij}}{M_{\text{pl}}} }^2 4 (k_1\cdot k_2)\,,
\end{equation}
with which corresponding decay rate is shown to be
\begin{align}
\Gamma(\phi \to \widetilde{N}^c_i  (L_j \cdot \widetilde{H}) ) & = 2\times \left|\frac{\lambda_{3}^{ij}}{M_{\text{pl}}} \right|^2 \frac{m_\phi^3}{768 \pi ^3} \left[ 1+9 \mu_N-9 \mu_N^2-\mu_N^3+6 \mu_N \log(\mu_N)+6 \mu_N^2 \log(\mu_N)
\right]\,.
\end{align}
For $\phi(p) \to \widetilde{H}(k_1) \widetilde{L}(k_2) N^c(k_3)$ (Fig.~\ref{fig:3-body3}), the squared matrix element is 
\begin{equation}
\abs{\mathcal{M}}^2=\abs{\frac{\lambda_{3}^{ij}}{M_{\text{pl}}} }^2 4 (k_1\cdot k_3)\,,
\end{equation}
and decay rate
\begin{align}
\Gamma(\phi \to  N^c_i(\widetilde{L}_j \cdot \widetilde{H}))&= 2\times \left|\frac{\lambda_{3}^{ij}}{M_{\text{pl}}} \right|^2  \frac{m_\phi^3}{768 \pi ^3} \left[1-6 \mu_N+3 \mu_N^2+2 \mu_N^3-6 \mu_N^2 \log(\mu_N)\right]\,.
\end{align}
Finally, for the inflaton decays into three scalars  $\phi(p) \to H(k_1) \widetilde{L}(k_2) \widetilde{N}^c(k_3)$ (Fig.~\ref{fig:3-body4}), the squared matrix element is 
\begin{equation}
\abs{\mathcal{M}}^2=\left|\frac{\lambda_4^{ij} \Lambda_N}{M_{\text{pl}}} \right|^2\,,
\end{equation}
and decay rate reads
\begin{align}
\Gamma(\phi \to \widetilde{N}^c_i (\widetilde{L}_j \cdot H) ) & = 2\times \left|\frac{\lambda_4^{ij}  \Lambda_N}{M_{\text{pl}}} \right|^2 \times \frac{m_\phi}{512 \pi ^3} \left[ 1 -\mu_N^2+2 \mu_N \log(\mu_N)\right]\,.
\end{align}

%%%%%%%%%%%%%%%%%%%%%%%%%%%%%%%%%%%%%%%%%%%%%%%%%%

\section{\label{app:leptogenesis}Baryon asymmetry from non-thermal leptogenesis}

In this section, we discuss baryogenesis via leptogenesis \cite{Buchmuller:2004nz, Fong:2012buy}. There are two possible scenarios depending on the relative magnitudes of the reheating temperature, $\Trh$, and the right-handed neutrino masses. If the reheating temperature is high enough for the thermal production of right-handed neutrinos to be efficient, the subsequent out-of-equilibrium decay of these neutrinos can generate a baryon asymmetry through the sphaleron process. This mechanism is known as thermal leptogenesis~\cite{Fukugita:1986hr}. In thermal leptogenesis, inverse processes act as washout effects that suppress the resulting asymmetry. Consequently, thermal leptogenesis typically requires a high reheating temperature, which can lead to the gravitino problem~\cite{Khlopov:1984pf, Ellis:1984eq, Kawasaki:1994af}. On the other hand, if the reheating temperature is low, the thermal production of right-handed neutrinos will be Boltzmann suppressed. However, it has been noted that the inflaton’s non-thermal two-body decay into pairs of right-handed neutrinos can still account for the baryon asymmetry of the universe~\cite{Asaka:1999yd, Asaka:1999jb,Fujii:2002jw}. More recently, it was shown that the inflaton’s non-thermal three-body decay can also successfully lead to leptogenesis~\cite{Drees:2024hok}.

For baryogenesis via leptogenesis, it is typically required that the reheating temperature be higher than the electroweak scale to ensure the sphaleron process is efficient. In the current inflationary setup, the inflaton mass has been shown to be smaller than $\mathcal{O}(10^8)~\textrm{GeV}$, as discussed at the end of Sec.~\ref{sec:modular_inflation}. Consequently, the reheating temperature remains below $\mathcal{O}(100)\textrm{GeV}$, assuming the inflaton decays into neutrino channels (cf. Fig.~\ref{fig:Trh}). Nevertheless, given the novel feature of the current lepton flavor model, which not only resolves the lepton flavor puzzle but also naturally provides channels for reheating, it remains interesting to investigate the lower bound on the inflaton mass that would lead to the observed baryon asymmetry of the universe (BAU). To this end, we treat the inflaton mass as a free parameter.

As discussed in the previous section, in our scenario reheating temperature is lower than  the lightest right-handed 
neutrino mass, which implies that the thermal leptogenesis is suppressed in our scenario.  In this work, we will focus on the non-thermal case, the produced baryon asymmetry from right handed neutrino decay can be estimated as~\cite{Asaka:1999yd,Drees:2024hok}:
\begin{align}
Y_{B}\equiv \frac{n_B}{s} \simeq -\frac{8}{23} \times \frac{3}{4}\frac{T_{\text{rh}}}{m_\phi} \sum_i \epsilon_i \times \left[2\text{Br}\big(\phi\rightarrow {N}_i+{N}_i\big)+\text{Br}\big(\phi\rightarrow {N}_i+\text{others}\big)\right]\,,
\label{eq:baryon_asymmetry}
\end{align}
where $i$ sums over all the right handed neutrinos produced from inflaton decays. The first factor $-8/23$ is the conversion factor which transfer lepton asymmetry to baryon asymmetry~\cite{Khlebnikov:1988sr,Harvey:1990qw}.  The $\epsilon_i$ measures the asymmetry in the right handed neutrino decays:
\begin{equation}\label{eq:epsilon}
\epsilon_i = \frac{\Gamma (N_i \to H_u+L)-\Gamma(N_i \to \overline{H}_u+\overline{L})}{\Gamma(N_i \to H_u+L)+\Gamma(N_i \to \overline{H}_u+\overline{L})}\,,
\end{equation}
where the decay process should also include SUSY channels. i.e. $N_i \to \tilde{H}_u+\tilde{L}$. In our model, we have two semi-degenerate right handed neutrinos $M_i=\Lambda_N (1.372,1.447,2.818)$. This leads to an enhancement of $\epsilon_i$, which should be evaluated as~\cite{Pilaftsis:2003gt}:
\begin{equation}
    \epsilon_{i} = \frac{\Im{( h h^\dagger)^2_{ij}}}{( h h^\dagger)_{ii}( h h^\dagger)_{jj}} \frac{(M_i^2-M_j^2)M_i~\Gamma_{N_j}}{(M_i^2-M_j^2)^2+M_i^2~\Gamma_{N_j}^2}\,, 
\end{equation}
where $i,j$ run over $1,2$ in our model. When $i=1$, one should take $j=2$ and vice versa. $h$ is the Yukawa coupling between right handed neutrino, lepton and higgs field. In the bases where right handed neutrinos are diagonal, it reads:
\begin{equation}
h=U_N^T {\cal Y}_D U_L^\nu=  \left(
\begin{array}{ccc}
 1.372i ~&~ -0.347 ~&~ 0.009i \\
 0.347 ~&~ 1.447i~&~ 0.001 \\
 0.009i ~&~ -0.001 ~&~ -2.818i \\
\end{array}
\right)g_1\,.
\end{equation}

$\Gamma_{N_i}$ is the decay width of right handed neutrinos. At tree level, it reads:
\begin{equation}
    \Gamma_{N_i} = \frac{(h h^\dagger)_{ii}}{4\pi}M_i
\label{eq:Ndecay}\,.
\end{equation}
We note that the decay of sneutrinos can also generate a CP asymmetry, analogous to eq.~\eqref{eq:epsilon}. However, their contribution to the BAU is small due to the domination of the branching ratio into heavy neutrinos from inflaton decays (cf. eq.~\eqref{eq:phiNN} and eq.~\eqref{eq:phiNtildeNtilde}).
The BAU at present is given by
\cite{Kolb:1990vq}
\begin{equation}\label{etaB}
\eta_B \equiv \frac {n_B} {n_\gamma} = \left( \frac {s} {n_\gamma} \right)_0
\left( \frac {n_B} {s} \right)
\simeq 7.02 \times Y_{B}\,,
\end{equation}
where $ n_\gamma = \frac{2\zeta(3)T^3}{\pi^2} $ is the photon number density and $s=\frac{2\pi^2}{45} g_{\star\,s} T^3$ corresponding to the entropy density. The subscript $ 0 $ refers to the current time, where $ T = T_0 \simeq 2.73 $~K and $ g_{\star\,s} \simeq 3.9$. Using the baryon asymmetry of the Universe (BAU) value based on Planck 2018 \cite{Fields:2019pfx}, 
\begin{equation} \label{eq:YBL_required}
\eta^{\text{exp}}_B \simeq \left(6.143 \pm  0.190 \right)\times 10^{-10}\,,    
\end{equation}
we can obtain the required $Y_{B}$ to match the observation, which is $Y^{\text{exp}}_{B} =\frac{6.143}{7.02} \times \eta^{\text{exp}}_B \simeq 8.75 \times 10^{-11}$.
%%%%%%%%%%%%%%%%%%%%%%%%%%%%%%%%%%%%%%%%%%%%%%%%%%%
\subsection{Parameter space}
Now we have all the relevant ingredients to calculate the baryon asymmetry in this model.  
%%%%%%%%%%%%%%%%%%%%%%%%%%%%%%%%%%%%%%%%%%%%%%%%%%%%%%%
\begin{figure}[t!]
\centering
\includegraphics[width=0.43\textwidth]{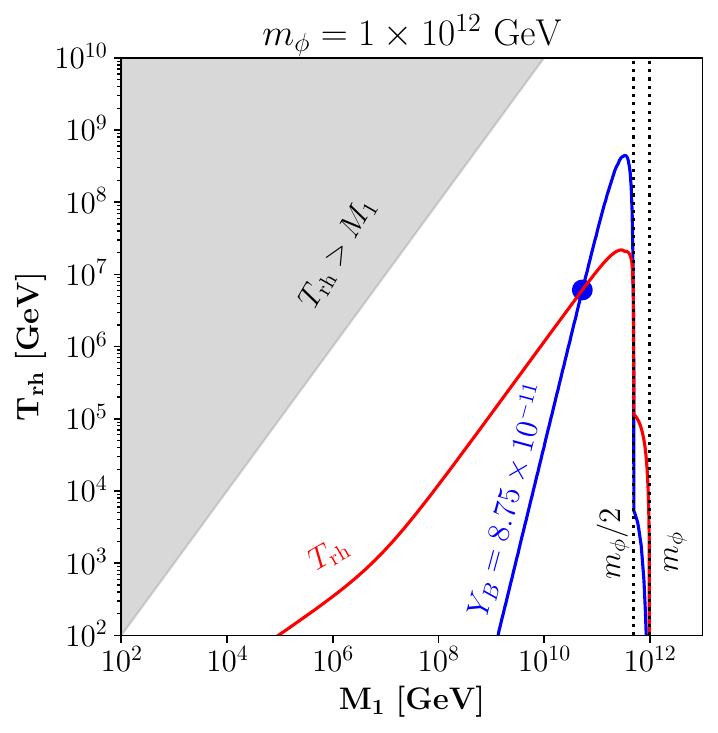}
\includegraphics[width=0.45\textwidth]{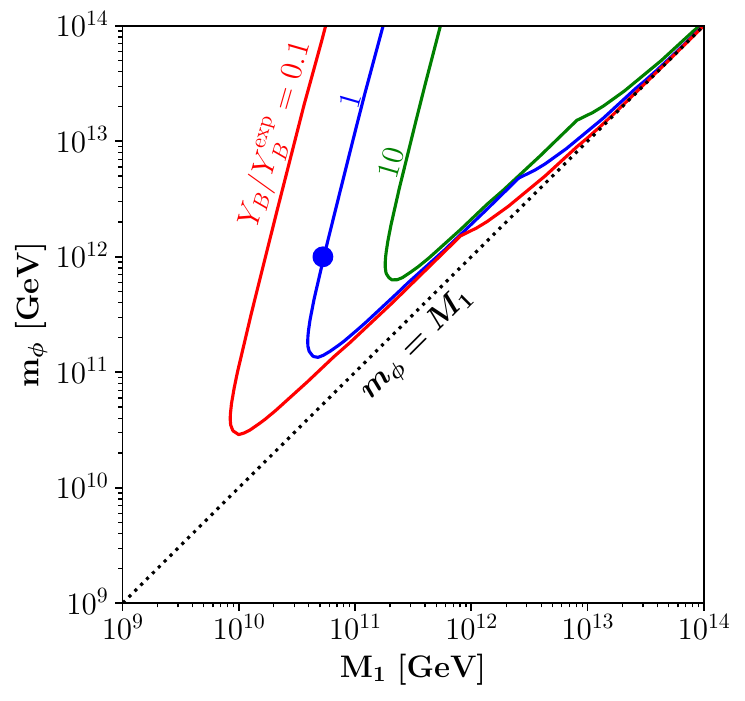}
\caption{Left panel: $\Trh$ as function of $M_1$ to yield $Y_B =8.75 \times 10^{-11}$ with $m_\phi =10^{12}~\text{GeV}$ (blue line) by considering $\Trh$ as a free parameter. The red line corresponds to the minimal reheating temperature in our scenario. Right panel: $(m_\phi, M_1)$ scan by assuming the minimum reheating scenario, i.e. the reheating channel also sources leptogenesis with $Y_B = 8.75 \times 10^{-12}$ (red), $Y_B = 8.75 \times 10^{-11}$ (blue) and $Y_B = 8.75 \times 10^{-10}$ (green).}
\label{fig:ps_BAU}
\end{figure}
%%%%%%%%%%%%%%%%%%%%%%%%%%%%%%%%%%%%%%%%%%%%%%%%%%%%%%%%%%%%%%%%%%%%%%%%%
The results are shown in Fig.~\ref{fig:ps_BAU}. As an example, in the left panel, we consider an inflaton mass of $m_\phi = 10^{12}~\text{GeV}$. The blue line represents the parameter space for $\Trh$ as a function of $M_1$, required to yield $Y_B = 8.75 \times 10^{-11}$. Note the branching ratios change with reheating temperature. To achieve this, we treat $\Trh$ as a free parameter. When considering inflaton two-body and three-body decays account for reheating, the corresponding reheating temperature is shown as the red line. It intersects the blue curve at $M_1 \simeq 5.3\times 10^{10}~\text{GeV}$ and $\Trh \simeq 6.1 \times 10^6~\text{GeV}$, as indicated by the blue dot, which represents the allowed parameter space in our scenario when $m_\phi = 10^{12}~\text{GeV}$. Varying the inflaton mass shifts the intersection point of the red and blue lines. Moreover, we note that as $M_1 \to m_\phi / 2$, the red line tends to merger with the blue curve due to the contribution of three-body decay to $Y_B$. In other words, in the regime where $M_1 < m_\phi / 2$, two-body decays dominate, and three-body decays take over when $m_\phi / 2 < M_1 < m_\phi$. This also explains the features of the blue curve between the two vertical black dotted lines. Finally, we note that $M_1 \gg \Trh$, validating the assumption of non-thermal leptogenesis.

In the right panel, we show a $  (m_\phi, M_1) $ scan that results in  $Y_B = 8.75 \times 10^{-12}$ (red), $Y_B = 8.75 \times 10^{-11}$ (blue) and $Y_B = 8.75 \times 10^{-10}$ (green), assuming the reheating channel also accounts for leptogenesis. We note that all the curves would approach to the black dotted line with $M_1 = m_\phi$ as explained and implied by the figure in the left panel. To explain the BAU observed in our universe, the allowed parameter space is indicated by the blue curve, with the blue dot corresponding to the same point as shown in the left panel.  For a fixed $ Y_B $, in the region where $ M_1 < m_\phi / 2 $, the inflaton mass scales as $ m_\phi \propto M_1^2 $, as shown in the right panel of Fig.~\ref{fig:ps_BAU}, due to the dominance of two-body decays. We find that a lower bound on the inflaton mass around $m_\phi \gtrsim 10^{11}\text{GeV}$ is required to explain the entirety of the observed baryon asymmetry, as shown in the edge of the blue line in  the right panel of Fig.~\ref{fig:ps_BAU}. This also implies the lightest right handed neutrino mass should satisfy $M_1 \gtrsim 10^{11}\text{GeV}$ to give rise to the observed BAU.

We note that the lower bounds mentioned above could be relaxed if we assume that the non-thermal leptogenesis under consideration accounts for only part of the observed baryon asymmetry. For instance, $m_\phi$ can be as small as $10^{10}\,\textrm{GeV}$ if we assume that non-thermal leptogenesis contributes only 10\% of the BAU, as demonstrated by the red line in the right panel of Fig.~\ref{fig:ps_BAU}. 

In the current small-field inflationary setup, the inflaton has a very small mass, $m_\phi \lesssim 10^8~\text{GeV}$, as shown at the end of section ~\ref{sec:modular_inflation}. This leads to a low reheating temperature, $\Trh \lesssim 100~\text{GeV}$, as illustrated in Fig.~\ref{fig:Trh}. In such a regime, it becomes very challenging to generate a sizable contribution to the observed BAU. The failure to reproduce the correct BAU in our model stems from two main factors. First, in our setup, the same channel responsible for reheating the Universe also produces right-handed neutrinos. When the reheating temperature is low, this results in a suppressed right-handed neutrino abundance. Second, for $\Trh \lesssim 100\text{GeV}$, electroweak sphaleron processes become inefficient, further reducing the conversion of lepton asymmetry into baryon asymmetry. These two features together render non-thermal leptogenesis ineffective in our small-field inflationary scenario. An alternative way forward is to investigate the possibility of raising the inflationary scale, such as through large field inflation \cite{Starobinsky:1980te, Kallosh:2013hoa, Kallosh:2013maa, Drees:2022aea}, which could give rise to inflaton mass as larger as $\mathcal{O}(10^{13})~\text{GeV}$. 

Recent developments in realizing large-field inflation within the modular invariant framework are discussed in Ref.~\cite{Casas:2024jbw} for Starobinsky inflation and Refs.~\cite{Kallosh:2024ymt, Kallosh:2024pat} for the $\alpha$-attractor scenario. In these frameworks, the reheating temperature could be higher due to an increased inflaton mass scale, potentially facilitating successful baryogenesis.

\end{appendix}

% \flushbottom
% \newpage
\bibliographystyle{JHEP}
\bibliography{ref}
\end{document}